\documentclass[prb,a4paper,final,superscriptaddress,longbibliography]{revtex4-1}
\usepackage{bm}
\usepackage{graphicx}
\usepackage{makeidx}
\usepackage{amsmath,amssymb}

\begin{document}


\title{$d_{x^2-y^2}$-wave density wave and $d_{x^2-y^2}$-wave superconducting gap on the extended Hubbard model on a square lattice}

	

\author{Yuhei Hirose$^{\ast}$\footnote[0]{$^{\ast}$Email:\;hiro3332@gmail.com}}
\altaffiliation{}
\affiliation{Department of Physics, Faculty of Science and Technology, Tokyo University of Science, Noda, Chiba 278-8510, Japan}
\author{Akihide Oguchi}
\altaffiliation{}
\affiliation{Department of Physics, Faculty of Science and Technology, Tokyo University of Science, Noda, Chiba 278-8510, Japan}
\author{Yoshiyuki Fukumoto$^{\dagger}$\footnote[0]{$^{\dagger}$Email:\;yfuku@rs.tus.ac.jp}}
\altaffiliation{}
\affiliation{Department of Physics, Faculty of Science and Technology, Tokyo University of Science, Noda, Chiba 278-8510, Japan}

\date{\today}

\begin{abstract}
The extended Hubbard model with a nearest-neighbor Coulomb repulsion on the square lattice is studied to obtain insight into the phase diagram of cuprate high $T_c$ superconductors (HTS). 
To pursue the hidden-order scenario proposed in [S. Chakravarty \textit{et al.}, Phys. Rev. B $\bm{63}$, 094503 (2001)], we derive an effective Hamiltonian by using the canonical transformation and develop a mean-field theory. 
The calculated phase diagrams are qualitatively consistent with the experimental phase diagrams of HTS, and we thus 
conclude that the pseudogap can be interpreted as the order parameter of the $d_{x^2-y^2}$-wave density wave (DDW) state, and the $d_{x^2-y^2}$-wave superconducting (DSC) rises based on the DDW order. 
Furthermore, the analytical representation of the density of states is obtained and, near the optimal doping of the DSC, the van Hove singular point of the density of states is located at the Fermi level. 
\end{abstract}

\pacs{}

\maketitle


\section{Introduction}

Since cuprate high $T_c$ superconductors (HTS) were discovered in 1986\cite{ref.HTS}, the mechanism of HTS has been one of the most unsolved issues in condensed matter physics.\cite{ref.Anderson1,ref.Anderson2,ref.Nagaosa1,ref.Anderson3,ref.Anderson4}
In this study, we examine the square-lattice extended Hubbard Hamiltonian with the nearest-neighbor (NN) Coulomb repulsion to obtain the physical properties of HTS. 
The reasons for introducing this Hamiltonian are as follows. 

First, many published papers on HTS are based on the Hubbard Hamiltonian.\cite{ref.Anderson2,ref.Nagaosa1,ref.Anderson3,ref.Anderson4}
The reason for this may be that the mother substances of HTS show the antiferromagnetic (AF) ground state, which is described by the Hubbard Hamiltonian with on-site large Coulomb repulsion $U$. 
In HTS, many theories on underdoped regions have been developed from the spin-spin interactions obtained from the Hubbard Hamiltonian by using the canonical transformation.\cite{ref.Anderson2,ref.Nagaosa1,ref.Anderson3,ref.Anderson4}
However, the stripe states and, recently, the charge density wave (CDW) states, which contain two-electron-occupied sites, have been identified in the pseudogap phase of many HTS.\cite{ref.J.E.Hoffman,ref.M.Vershinin,ref.E.H.da,ref.K.Fujita,ref.G.Ghiringhelli,ref.J.Chang1,ref.A.J.Achkar1,ref.W.Tabis,ref.R.Comin1,ref.R.Comin2,ref.S.Gerber,ref.J.Chang2,ref.Y.Y.Peng,ref.T.Wu1,ref.T.Wu2,ref.T.Wu3,ref.kawasaki2015,ref.kawasaki}
These have been observed in x-ray scattering,\cite{ref.E.H.da,ref.G.Ghiringhelli,ref.J.Chang1,ref.A.J.Achkar1,ref.W.Tabis,ref.R.Comin1,ref.R.Comin2,ref.S.Gerber,ref.J.Chang2,ref.Y.Y.Peng} tunneling microscopy,\cite{ref.J.E.Hoffman,ref.R.Comin1,ref.M.Vershinin,ref.E.H.da,ref.K.Fujita} nuclear magnetic resonance,\cite{ref.T.Wu1,ref.T.Wu2,ref.T.Wu3,ref.kawasaki2015,ref.kawasaki} and other experiments. 
We can hardly explain these experimental results by using the Hubbard Hamiltonian with a large $U$. 
Therefore, we invoke the model Hamiltonian which includes the NN Coulomb repulsion $V$ in addition to the on-site Coulomb repulsion $U$.\cite{ref.Y.Ohta,ref.Onozawa,ref.Nayak,ref.Fukumoto1,ref.Fukumoto2,ref.Fukumoto3,ref.Aoki1,ref.Aoki2,ref.Aoki3,ref.Ogata}
For the half-filled case of this Hamiltonian, we have two types of regions depending on the magnitude of $U/V$. In this study, we call the region for $U>4V$ ($U<4V$) an AF (CDW) region. 

The second reason is as follows. The pseudogap phase,\cite{ref.Pseudogap1} which have been one of the most incomprehensible parts in the phase diagrams of HTS,\cite{ref.Nagaosa1,ref.Pseudogap2,ref.Pseudogap4,ref.Pseudogap7,ref.Pseudogap10} is observed in the phase diagrams of HTS, and the $d_{x^2-y^2}$-wave density wave (DDW) has been considered one of the strong candidates for explaining the pseudogap phase.\cite{ref.Nayak,ref.Chakravarty,ref.Tewari,ref.Chakravarty2,ref.Laughlin1,ref.Laughlin2,ref.Makhfudz}
S. Chakravarty \textit{et al.} proposed that this pseudogap is characterized by a hidden broken symmetry of $d_{x^2-y^2}$-type, which is DDW order, and is an actual gap in the one particle excitation spectrum.\cite{ref.Chakravarty}
S. Chakravarty \textit{et al.} also indicated that the point-contact tunneling measurement shows the considerable size of the tunneling gap, the $d_{x^2-y^2}$-wave-like gap persists above the superconducting $T_c$, and many other experiments are consistent with the presence of DDW.\cite{ref.Chakravarty,ref.Ch.Renner}
Therefore, considering the aforementioned, we assume that the pseudogap derives from the DDW order. Finally, we derive the effective Hamiltonian using the canonical transformation to discuss the $d_{x^2-y^2}$-wave superconducting (DSC) condensation in DDW metal. 

Previously, based on the CDW region on the extended Hubbard Hamiltonian with NN Coulomb repulsion $V$ and using the canonical transformation, when the hole is doped and the system deviates from the half filling, we found that $d_{xy}$-wave superconductivity coexists with the CDW order.\cite{ref.Onozawa} In this study, considering the AF region, we investigate whether DSC condensation occurs even in DDW instead of CDW metal. Our obtained phase diagrams prove to be qualitatively consistent with the experimental phase diagrams of HTS\cite{ref.experiment} and show that a metal-metal quantum transition point exists under the DSC dome, which had been proposed by S. Chakravarty \textit{et al.}\cite{ref.Chakravarty}

The remainder of this paper is organized as follows. 
In Sect.\;II, the extended Hubbard Hamiltonian is defined and the mean-field effective Hamiltonian is derived using the canonical transformation. 
In Sect.\;III, we introduce the DDW order parameters and consider the DDW phase in the underdoped region. 
In addition, we obtain the self-consistent equations for the DDW order parameters. 
In Sect.\;IV, we study the DSC phase by introducing the DSC order parameters, and then derive the self-consistent equations for these parameters. 
In Sect.\;V, we obtain the phase diagrams by numerically solving the derived self-consistent equations. 
Furthermore, we obtain the Fermi surface and density of states, and then discuss the physical properties of the DDW state. 
In Sect.\;VI, we summarize the results of this study.

\section{Hamiltonian}

We begin with the Hamiltonian, defined as: 
\begin{equation}
H_0=U\sum_{i}n_{i,\uparrow}n_{i,\downarrow}+V\sum_{\langle i,j\rangle}n_i n_j, 
\label{1} 
\end{equation}
where $n_{i,\sigma}=c_{i,\sigma}^{\dagger}c_{i,\sigma}$ $(\sigma=\uparrow, \downarrow)$ is the electron number operator at the $i$th site, $c_{i,\sigma}^{\dagger}$ $(c_{i,\sigma})$ is the electron creation (annihilation) operator, and $n_i=n_{i,\uparrow}+n_{i,\downarrow}$. The summation runs over the square lattice and $\langle i,j\rangle$ is the NN pair of sites. We divide the square lattice into the A and B sublattices. 

\begin{figure}[t]
\begin{center}
\includegraphics[width=.70\linewidth]{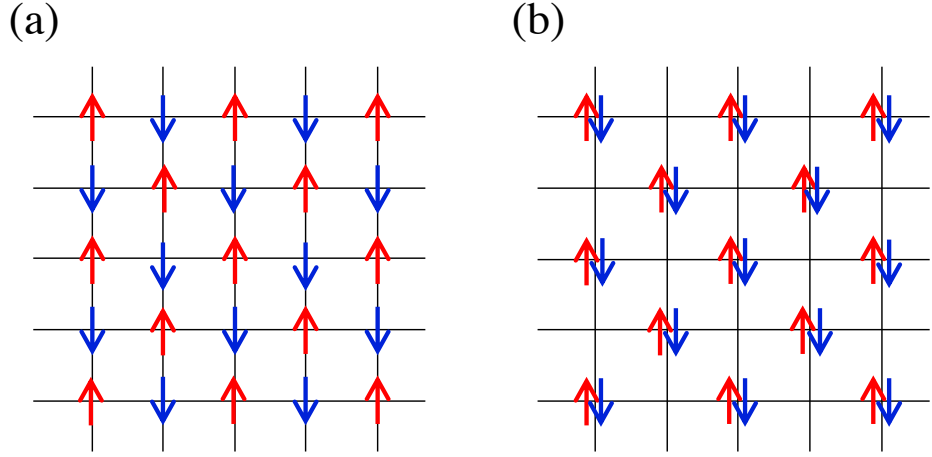}
\end{center}
\caption{Distribution of electrons in the (a) AF and (b) CDW regions for the half-filled case. The arrows represent the electrons whose spin direction is that of the arrow. The sublattice composed from (a) up spins for the AF region and (b) two-electrons-occupied sites for the CDW region is called the A sublattice. By contrast, the sublattice composed from (a) down spins for the AF region and (b) no occupied sites for the CDW region is called the B sublattices.}
\label{fig:1}
\end{figure}

For the half-filled case, we have two types of regions. One is the AF region, which is composed of up and down spins in the A and B sublattices, respectively, as shown in FIG.~\ref{fig:1}(a). 
Another is the CDW region, which is composed of two-electrons-occupied sites in the A sublattice and vacant sites in the B sublattice, as shown in FIG.~\ref{fig:1}(b). 
Because we have $H_0|\rm{AF}\rangle=$$2NV$$|\rm{AF}\rangle$ and $H_0|\rm{CDW}\rangle=$$\frac{N}{2}$$|\rm{CDW}\rangle$, where $N$ is the number of sublattice sites, we have an AF (CDW) region in the case of $U>4V$ ($U<4V$). 
The mother substances of HTS show an AF ground state. Therefore, in this study, we consider the AF region of $U>4V$.

As soon as holes are doped in the mother substances, the electrons begin to transfer from site to site. 
Therefore, we introduce the Hamiltonian of the transition of the electrons as: 
\begin{align}
H_1&=-t\sum_{i,\delta,\sigma}(c_{i,\sigma}^{\dagger}c_{i+\delta,\sigma}+\rm{h.c.}) 
\label{2},
\end{align}
where $\delta=\pm\bm{e}_{x}, \pm\bm{e}_{y}$ are NN vectors, and we assume that $t\ll U$ throughout this study.

To describe the AF region explicitly, we introduce hole creation and annihilation operators and then rewrite $c_{i,\uparrow}^{\dagger} \rightarrow b_{i}$, $c_{i,\downarrow}\rightarrow a_i $ for $i\in$ A sublattice and $c_{l,\downarrow}^{\dagger} \rightarrow d_{l}$, $c_{l,\uparrow} \rightarrow c_l$ for $l \in$ B sublattice. 
Hereafter, for convenience, we refer to these notations as hole representations.
Substituting the hole representation into Eqs.\;(\ref{1}) and (\ref{2}), we obtain
\begin{align}
H_0=&2NV+\sum_{i \in \rm{A}}\left\{(U+4V)n_i^a-4V n_i^b-U n_i^an_i^b\right\}+\sum_{l\in \rm{B}}\left\{(U+4V)n_l^c-4V n_l^d-U n_l^c n_l^d\right\} \notag \\
&+V\sum_{i \in \rm{A}}\sum_{\delta}(n_i^a-n_i^b)(n_{i+\delta}^c-n_{i+\delta}^d), 
\label{3} 
\end{align}
and
\begin{align}
H_1&=-t\sum_{i,\delta}\left\{(b_ic_{i+\delta}+a_i^{\dagger}d_{i+\delta}^{\dagger})+\textrm{h.c}\right\}
\label{4},
\end{align}
where $n_i^a=a_i^{\dagger}a_i$, $n_i^b=b_i^{\dagger}b_i$, $n_l^c=c_l^{\dagger}c_l$, and $n_l^d=d_l^{\dagger}d_l$. 

Here, we divide $H_1$ into two parts written as:
\begin{align}
H_1=&H_1(2,0)+H_1(1),
\label{5} 
\end{align}
with
\begin{align}
H_1(2,0)=&-t\sum_{i,\delta}\Bigl[\bigl(b_{i}c_{i+\delta}+c_{i+\delta}^{\dagger}b_{i}^{\dagger}\bigr)\left\{(1-n_i^a)n_{i+\delta}^d+n_i^a(1-n_{i+\delta}^d)\right\}\notag \\
&\qquad\quad+\bigl(a_{i}^{\dagger}d_{i+\delta}^{\dagger}+d_{i+\delta}a_{i}\bigr)\left\{n_i^b(1-n_{i+\delta}^c)+(1-n_i^b)n_{i+\delta}^c\right\}\Bigr]
,\label{6}\\
H_1(1)=&-t\sum_{i,\delta}\Bigl[\bigl(b_{i}c_{i+\delta}+c_{i+\delta}^{\dagger}b_{i}^{\dagger}\bigr)\left\{n_i^an_{i+\delta}^d+(1-n_i^a)(1-n_{i+\delta}^d)\right\}\notag \\
&\qquad\quad+\bigl(a_{i}^{\dagger}d_{i+\delta}^{\dagger}+d_{i+\delta}a_{i}\bigr)\left\{(1-n_i^b)(1-n_{i+\delta}^c)+n_i^bn_{i+\delta}^c\right\}\Bigr]
\label{7},
\end{align}
where $H_1(2,0)|\rm{AF}\rangle=0$ is satisfied and $H_1(2,0)$ cannot perturb the AF region. By contrast, $H_1(1)|\rm{AF}\rangle\ne 0$ is satisfied and $H_1(1)$ can change the AF region.
Note that the numbers in parentheses in Eqs.\;(\ref{6}) and (\ref{7}) represent the number of holes per site in each region (i.e., (2,0) [(1)] represent the number of holes per site is two or zero [one]), which correspond to the CDW [AF] region.

Next, by using the canonical transformation, we obtain the effective Hamiltonian: 
\begin{align}
\mathcal{H}=&e^S (H_0+H_1) e^{-S}\notag \\
		 =&H_0+H_1(2,0)+H_1(1)+\left[S, H_0\right]+\left[S, H_1(2,0)\right]+\left[S, H_1(1)\right]\notag \\
		 &+\frac{1}{2}\bigl(\left[S, \left[S, H_0\right]\right]+\left[S, \left[S, H_1(2,0)\right]\right]+\left[S, \left[S, H_1(1)\right]\right]\bigr)+\cdots
\label{8},
\end{align}
where $S$ satisfies
\begin{gather}
H_1(1)+[ S, H_0 ] =0
\label{9}.
\end{gather}
From Eq.\;(\ref{9}), we obtain:
\begin{eqnarray}
S&=&-i\int_0^{\infty}d\lambda e^{i\lambda H_0}H_1(1)e^{-i\lambda H_0}
\label{10}.
\end{eqnarray}

We obtain $S$ in the hole representation by neglecting those terms consisting of the product of more than four Fermi operators, such as $n_i^an_i^bn_l^c$ and $a_ib_in_l^cn_l^d$, because $n$ is a small quantity in the underdoped region (see Appendix A). Finally, $S$ is written as:
\begin{align}
S=&\sum_{i,\delta}\Bigl\{I_0(1-N_{i,i+\delta}^{ad})+I_1\sum_{\eta(\ne\delta)}N_{i+\eta,i+\delta-\eta}^{cb}+I_2\sum_{\eta(\ne \delta)}N_{i+\eta,i+\delta-\eta}^{da}\Bigr\}F_{i,i+\delta}^{bc} \notag \\
&-\sum_{i,\delta}\Bigl\{I_0(1-N_{i,i+\delta}^{bc})+I_1\sum_{\eta(\ne\delta)}N_{i+\eta,i+\delta-\eta}^{d a}
+I_2\sum_{\eta(\ne \delta)}N_{i+\eta,i+\delta-\eta}^{c b}\Bigr\}F_{i,i+\delta}^{ad}
\label{11},
\end{align}
where
\begin{align}
I_0=&\dfrac{t}{U-V}
\label{12}, \\
I_1=&\dfrac{t}{U-2V}-\dfrac{t}{U-V}
\label{13},\\
I_2=&\dfrac{t}{U}-\dfrac{t}{U-V}
\label{14},\\
N_{i,i+\delta}^{ad}=&n_i^a+n_{i+\delta}^d
\label{15}, \\
N_{i+\delta-\eta,i+\eta}^{bc}=&n_{i+\delta-\eta}^b+n_{i+\eta}^c
\label{16},\\
N_{i+\delta-\eta,i+\eta}^{ad}=&n_{i+\delta-\eta}^a+n_{i+\eta}^d
\label{17}, \\
F_{i,i+\delta}^{bc}=&b_ic_{i+\delta}-c_{i+\delta}^{\dagger}b_i^{\dagger} 
\label{18}, \\
F_{i,i+\delta}^{ad}=&a_id_{i+\delta}-d_{i+\delta}^{\dagger}a_i^{\dagger}
\label{19}.
\end{align}
Therefore, substituting Eq.\;(\ref{11}) into Eq.\;(\ref{8}), the effective Hamiltonian is written as:
\begin{equation}
\mathcal{H}_{eff}=H_0+H_1+[S, H_0]+[S, H_1]
\label{20},
\end{equation}
which is adopted as an equation up to the first order of $S$. In addition, the other terms, which are a second or greater order of $S$, are omitted because these consist of the product of more than four operators.
We calculate Eq.\;(\ref{20}) by using the approximations (I) and (II), described as follows. 

(I) We neglect the terms consisting of the product of more than four operators, which were already used in the process of deriving Eq.\;(\ref{11}). 

(II) The hopping terms that are farther than the NN lattice points, such as $a_i^{\dagger}a_{i+\bm{e}_{x}+\bm{e}_{y}}$, are neglected because we expect that the short-range interactions yield more essential effects compared with the long-range interactions. 
For the same reason, next NN interactions, such as $n_i^a n_{i+\delta+\eta}^a$, and the terms with three or more different types of suffixes, such as $b_i c_{i+\delta}n_{i+\eta}^d$ and $a_ib_id_{j}c_l$, are all neglected. Therefore, we consider the interactions only between the NN lattice points.

We obtain the effective Hamiltonian after the straightforward calculations (see Appendix B). 
\begin{align}
\mathcal{H}_{eff}=&C+\mathcal{H}_t+\mathcal{H}_n +\mathcal{H}_{nn}+\mathcal{H}_{int},
\label{21}
\end{align}
\begin{align}
C=&2NV-8N t I_0
\label{22}, \\
\mathcal{H}_t=&-t\sum_{i,\delta}\left\{\bigl(n_i^a+n_{i+\delta}^d\bigr)\bigl(b_ic_{i+\delta}+c_{i+\delta}^{\dagger}b_i^{\dagger}\bigr) 
-\bigl(n_i^b+n_{i+\delta}^c\bigr)\bigl(a_id_{i+\delta}+d_{i+\delta}^{\dagger}a_i^{\dagger}\bigr)\right\}
\label{23}, \\
\mathcal{H}_n=&\mu\biggl(\sum_i n_i^a+\sum_l n_l^c\biggr)+\mu^{\prime}\biggl(\sum_i n_i^b+\sum_l n_l^d\biggr)
\label{24}, \\
\mathcal{H}_{nn}=&-U^{\prime}\biggl(\sum_i n_i^a n_i^b+\sum_l n_l^c n_l^d\biggr)+V^{\prime}\sum_{i,\delta}\bigl(n_i^b n_{i+\delta}^d+n_i^a n_{i+\delta}^c\bigr)-V^{\prime \prime}\sum_{i,\delta}\bigl(n_i^b n_{i+\delta}^c+n_i^an_{i+\delta}^d\bigr)
\label{25},  \\ 
\mathcal{H}_{int}=&4tI_0\sum_{i,\delta}F_{i,i+\delta}^{ad}F_{i,i+\delta}^{b c}
\label{26},
\end{align}
where
\begin{align}
\mu=&U+4V+8t(2I_0-3I_1-3I_2)
\label{27}, \\
\mu^{\prime}=&-4V+8t(2I_0-3I_1-3I_2)
\label{28}, \\
U^{\prime}=&U+16tI_0
\label{29}, \\
V^{\prime}=&V-4tI_0+12t I_2
\label{30}, \\ 
V^{\prime\prime}=&V-12tI_1
\label{31}, \\ 
F_{i,i+\delta}^{b c}=&b_ic_{i+\delta}-c_{i+\delta}^{\dagger}b_i^{\dagger}
\label{32}, \\
F_{i,i+\delta}^{a,d}=&a_id_{i+\delta}-d_{i+\delta}^{\dagger}a_i^{\dagger}
\label{33}.
\end{align}

\section{$d_{x^2-y^2}$-wave density wave phase in the underdoped region}

The experimental results show that, as holes are being doped in the mother substances, the AF phase disappears and the pseudogap phase emerges in the underdoped region. 
Furthermore, as the amount of doping increases, the DSC phase is emerging.\cite{ref.experiment}  
However, the pseudogap phase is always found in the DSC phase, and the DSC gap evolves continuously into the d-wave-like pseudogap without collapsing.\cite{ref.Chakravarty} 
This fact suggests that the DDW always exists behind the underdoped region as if it were shadow. 
In this section, we first consider the DDW phase without the DSC phase. 

\begin{figure}[t]
\begin{center}
\includegraphics[width=.70\linewidth]{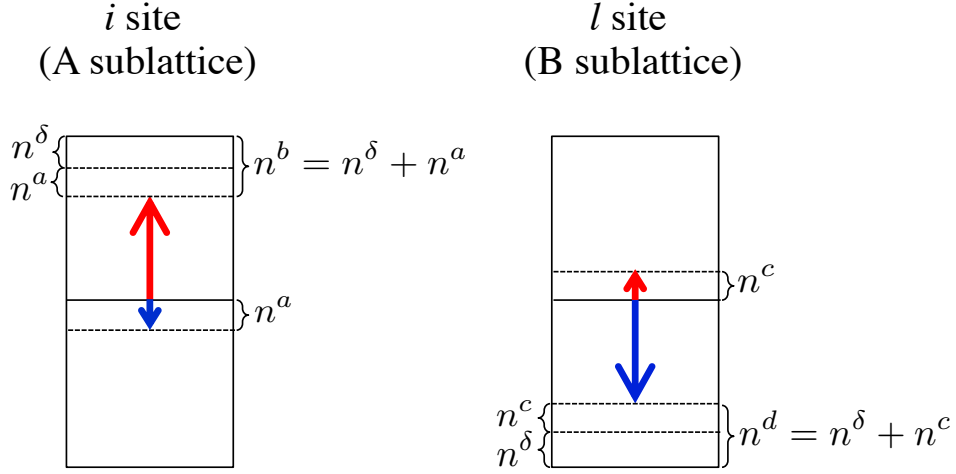}
\end{center}
\caption{Schematic representation of the electron configuration in $i$ and $l$ sites when the hole whose concentration is $n^{\delta}$ is doped at each site.}
\label{fig:2}
\end{figure}

We assume that the hole whose concentration is $n^{\delta}$ is doped at each site, and the hole concentrations with up and down spins are equivalent because of the use of symmetry. 
As shown in FIG.~\ref{fig:2}, at the $i$ site, the electron (hole) concentration with down (up) spin becomes $n^a$ $(n^{b}=n^{\delta}+n^a)$. 
Similarly, at the $l$ site, the electron (hole) concentration with up (down) spin becomes $n^c$ $(n^{d}=n^{\delta}+n^c)$, $n^a=n^c$, and $n^b=n^d$, where $n^a\equiv \langle n_i^a\rangle$, $n^b\equiv \langle n_i^b\rangle$, $n^c\equiv \langle n_l^c\rangle$, and $n^d\equiv \langle n_l^d\rangle$. 
Note that we use the symmetry whereby reversing the top and bottom of the $i$ site makes it equivalent to the $l$ site.
These electrons create the staggered magnetization at each site and we define the magnetization per site in the A sublattice as: 
\begin{align}
m=1-n^b-n^a=1-n^{\delta}-2n^a
\label{34},
\end{align}
which is the number of electrons with up spins minus that of electrons with down spins.
To conserve the doped concentration at each site, $n^{\delta}=n^b-n^a=n^d-n^c$, we introduce the chemical potentials $\mu_1$ and $\mu_2$, and write:
\begin{equation}
\mathcal{H}_n^{\prime}=\mathcal{H}_n+\mu_1\Bigl(\sum_i n_i^b-\sum_i n_i^a\Bigr)+\mu_2\Bigl(\sum_l n_l^d-\sum_l n_l^c\Bigr)
\label{35}.
\end{equation} 

Next, the mean-field approximations for the fourth-order terms in Eqs.\;(\ref{23}), (\ref{25}), and (\ref{26}) are adopted, and some of these are written as follows: 
\begin{align}
n_i^a b_ic_{i+\delta}& \Longrightarrow n^ab_ic_{i+\delta}+n_i^a\langle b_ic_{i+\delta}\rangle-n^a\langle b_ic_{i+\delta}\rangle
\label{36},\\
n_i^b a_id_{i+\delta}& \Longrightarrow n^ba_id_{i+\delta}+n_i^b\langle a_id_{i+\delta}\rangle-n^b\langle a_id_{i+\delta}\rangle
\label{37},
\end{align}
where the quantities $\langle b_ic_{i+\delta}\rangle$ and $\langle a_id_{i+\delta}\rangle$ can be represented by $\langle c_{i,\uparrow}^{\dagger} c_{i+\delta,\uparrow}\rangle$ and $ \langle c_{i,\downarrow}c_{i+\delta,\downarrow}^{\dagger}\rangle$, respectively, in the electron representations. These quantities refer to the charge densities. 
Furthermore, we omit terms such as $\langle d_{i+\delta}b_i \rangle (=\langle c_{i+\delta,\downarrow}^{\dagger}c_{i,\uparrow}^{\dagger}\rangle)$ and $\langle c_{i+\delta}^{\dagger}a_i^{\dagger}\rangle (=\langle c_{i+\delta,\uparrow}^{\dagger}c_{i,\downarrow}^{\dagger}\rangle)$, which give rise to the Cooper pair of holes, and terms such as $\langle a_{i}^{\dagger}b_i \rangle (=\langle c_{i,\downarrow}^{\dagger}c_{i,\uparrow}^{\dagger}\rangle)$ and $\langle d_{i+\delta}c_{i+\delta}^{\dagger}\rangle (=\langle c_{i+\delta,\downarrow}^{\dagger}c_{i+\delta,\uparrow}^{\dagger}\rangle)$ because we consider the large Coulomb repulsion $U$. 
The point contact tunneling experiment shows that the $d_{x^2-y^2}$-wave-like tunneling gap is observed in the normal state above the superconducting state.\cite{ref.Ch.Renner} 
Therefore, we introduce the DDW order parameters as:
\begin{align}
\langle a_i d_{i+\delta} \rangle =\begin{cases}
 \langle a_i d_{i\pm \bm{e}_x} \rangle = \Delta \\
 \langle a_i d_{i\pm \bm{e}_y} \rangle =- \Delta
\end{cases}
\label{38},
\end{align}
and
\begin{align}
\langle b_i c_{i+\delta} \rangle =\begin{cases}
 \langle b_i c_{i\pm \bm{e}_x} \rangle = \Delta^{\prime} \\
 \langle b_i c_{i\pm \bm{e}_y} \rangle =- \Delta^{\prime}
\end{cases}
\label{39}.
\end{align}

By using the mean-field approximation, we obtain the mean-field Hamiltonian of Eqs.\;(\ref{23}), (\ref{25}), and (\ref{26}) as:
\begin{align}
\mathcal{H}_{t}^{\rm{MF}}=&-t(1-m)\sum_{i,\delta}\left\{\bigl(b_i c_{i+\delta}+c_{i+\delta}^{\dagger}b_i^{\dagger}\bigr)-\bigl(a_i d_{i+\delta}+b_{i+\delta}^{\dagger}a_i^{\dagger}\bigr)\right\}
\label{40},\\
\mathcal{H}_{nn}^{\rm{MF}}=&\Bigl\{4V^{\prime}n^a-\bigl(U^{\prime}+4V^{\prime\prime}\bigr)n^b\Bigr\}\biggl(\sum_{i}n_i^a+\sum_{l}n_l^c\biggr)+\Bigl\{4V^{\prime}n^b-\bigl(U^{\prime}+4V^{\prime\prime}\bigr)n^a\Bigr\}\biggl(\sum_{i}n_i^b+\sum_{l}n_l^d\biggr)\notag \\
&-V^{\prime\prime}\sum_{i}\Bigl\{\Delta^{\prime \dagger}\bigl(b_{i}c_{i+\bm{e}_x}+b_{i}c_{i-\bm{e}_x}-b_{i}c_{i+\bm{e}_y}-b_{i}c_{i-\bm{e}_y}\bigr)+\rm{h.c.}\notag \\
&\qquad\qquad+\Delta^{\dagger}\bigl(a_{i}d_{i+\bm{e}_x}+a_{i}d_{i-\bm{e}_x}-a_{i}d_{i+\bm{e}_y}-a_{i}d_{i-\bm{e}_y}\bigr)+\rm{h.c.}\Bigr\}\notag \\
&+N\biggl[U^{\prime}n^an^b-2V^{\prime}\Bigl\{{(n^a)}^2+{(n^b)}^2\Bigr\}+4V^{\prime\prime}n^an^b+2V^{\prime\prime}\bigl({|\Delta|}^2+{|\Delta^{\prime}|}^2\bigr)\biggr]
\label{41},\\
\mathcal{H}_{int}^{\rm{MF}}=&4tI_0\sum_{i}\Bigl\{\bigl(\Delta-\Delta^{\dagger}\bigr)\bigl(F_{i,i+\bm{e}_x}^{bc}+F_{i,i-\bm{e}_x}^{bc}-F_{i,i+\bm{e}_y}^{bc}-F_{i,i-\bm{e}_y}^{bc}\bigr)\notag \\
&\qquad\quad+\bigl(\Delta^{\prime}-\Delta^{\prime\dagger}\bigr)\bigl(F_{i,i+\bm{e}_x}^{ad}+F_{i,i-\bm{e}_x}^{ad}-F_{i,i+\bm{e}_y}^{ad}-F_{i,i-\bm{e}_y}^{ad}\bigr)\Bigl\}\notag \\
&-8tI_0N\bigl(\Delta-\Delta^{\dagger}\bigr)\bigl(\Delta^{\prime}-\Delta^{\prime\dagger}\bigr)
\label{42},
\end{align}
where we use $n^a=n^c$ and $n^b=n^d$. 
From Eqs.\;(\ref{35}) and (\ref{40})--(\ref{42}), we finally obtain the mean-field total Hamiltonian in the DDW phase:
\begin{align}
\mathcal{H}_{\rm{DDW}}^{\rm{MF}}=&C+\mathcal{H}_t^{\rm{MF}}+\mathcal{H}_n^{\prime}+\mathcal{H}_{nn}^{\rm{MF}}+\mathcal{H}_{int}^{\rm{MF}} \notag \\
=&\sum_i\bigl(\epsilon_a n_i^a +\epsilon_b n_i^b\bigr)+\sum_l\bigl(\epsilon_c n_l^c+\epsilon_d n_l^d\bigr)-t(1-m)\sum_{i,\delta}\left\{\bigl(b_i c_{i+\delta}+c_{i+\delta}^{\dagger}b_i^{\dagger}\bigr)-\bigl(a_i d_{i+\delta}+d_{i+\delta}^{\dagger}a_i^{\dagger}\bigr)\right\}\notag \\
&+\sum_i \Bigl[\left\{4tI_0(\Delta^{\prime}-\Delta^{\prime\dagger})-V^{\prime\prime}\Delta^{\dagger}\right\}\bigl(a_i d_{i+\bm{e}_x}+a_i d_{i-\bm{e}_x}-a_i d_{i+\bm{e}_y}-a_i d_{i-\bm{e}_y}\bigr)+\text{h.c.}\notag \\
&\quad\quad\;+\left\{4tI_0(\Delta-\Delta^{\dagger})-V^{\prime\prime}\Delta^{\prime\dagger}\right\}\bigl(b_i c_{i+\bm{e}_x}+b_i c_{i-\bm{e}_x}-b_i c_{i+\bm{e}_y}-b_i c_{i-\bm{e}_y}\bigr)+\text{h.c.} \Bigr] \notag \\
&+C+N\biggl[U^{\prime}n^an^b-2V^{\prime}\Bigl\{{(n^a)}^2+{(n^b)}^2\Bigr\}+4V^{\prime\prime}n^an^b-8tI_0\bigl(\Delta-\Delta^{\dagger}\bigr)\bigl(\Delta^{\prime}- \Delta^{\prime\dagger}\bigr)\notag \\
&\qquad\qquad+2V^{\prime\prime}\bigl(|\Delta|^2+|\Delta^{\prime}|^2\bigr)\biggr],
\label{43}
\end{align}
where
\begin{align}
\epsilon_a=&\mu-\mu_1-\bigl(U^{\prime} +4V^{\prime\prime}\bigr)n^b+4V^{\prime}n^a
\label{44},\\
\epsilon_b=&\mu^{\prime}+\mu_1-\bigl(U^{\prime}+4V^{\prime\prime}\bigr)n^a+4V^{\prime}n^b
\label{45},\\
\epsilon_c=&\mu-\mu_2-\bigl(U^{\prime} +4V^{\prime\prime}\bigr)n^b+4V^{\prime}n^a
\label{46},\\
\epsilon_d=&\mu^{\prime}+\mu_2-\bigl(U^{\prime}+4V^{\prime\prime}\bigr)n^a+4V^{\prime}n^b
\label{47}.
\end{align}
Furthermore, performing the Fourier transformation of $a_i$, $b_i$, $c_l$, and $d_l$:
\begin{align}
a_i=&\sqrt{\frac{2}{N}}\sum_{\bm{k}}a_{\bm{k}}e^{i\bm{k}\cdot\bm{r}_i}
\label{48},\\ 
b_i=&\sqrt{\frac{2}{N}}\sum_{\bm{k}}b_{\bm{k}}e^{i\bm{k}\cdot\bm{r}_i}
\label{49},\\ 
c_l=&\sqrt{\frac{2}{N}}\sum_{\bm{k}}c_{\bm{k}}e^{-i\bm{k}\cdot\bm{r}_l}
\label{50},\\ 
d_l=&\sqrt{\frac{2}{N}}\sum_{\bm{k}}d_{\bm{k}}e^{-i\bm{k}\cdot\bm{r}_l}
\label{51},
\end{align}
and using the symmetry relation $\epsilon_a=\epsilon_c$ and $\epsilon_b=\epsilon_d$, we obtain: 
\begin{align}
\mathcal{H}_{\rm{DDW}}^{\rm{MF}}=&C^{\prime}+\sum_{\bm{k}} \Bigl\{\epsilon_a\bigl(a_{\bm{k}}^{\dagger}a_{\bm{k}}+c_{\bm{k}}^{\dagger}c_{\bm{k}}\bigr)+\epsilon_b\bigl(b_{\bm{k}}^{\dagger}b_{\bm{k}}+d_{\bm{k}}^{\dagger}d_{\bm{k}}\bigr)+\Gamma_{\bm{k}}^{\dagger}a_{\bm{k}}d_{\bm{k}}+\Gamma_{\bm{k}}d_{\bm{k}}^{\dagger}a_{\bm{k}}^{\dagger}-\Gamma_{\bm{k}}^{\prime\dagger}b_{\bm{k}}c_{\bm{k}}-\Gamma_{\bm{k}}^{\prime}c_{\bm{k}}^{\dagger}b_{\bm{k}}^{\dagger}\Bigr\}
\label{52},
\end{align}
where
\begin{align}
\Gamma_{\bm{k}}=&t(1-m)\gamma_{\bm{k}}-s_{\bm{k}}\bigl\{V^{\prime\prime}\Delta+4tI_0\bigl(\Delta^{\prime}-\Delta^{\prime\dagger}\bigr)\bigr\}
\label{53},\\
\Gamma_{\bm{k}}^{\prime}=&t(1-m)\gamma_{\bm{k}}+s_{\bm{k}}\bigl\{V^{\prime\prime}\Delta^{\prime}+4tI_0\bigl(\Delta-\Delta^{\dagger}\bigr)\bigr\}
\label{54}, \\ 
\gamma_{\bm{k}}=&2(\cos k_x+\cos k_y)
\label{55},\\ 
s_{\bm{k}}=&2(\cos k_x-\cos k_y)
\label{56}, \\ 
\epsilon_a=&\epsilon_c=\mu-\mu_1-\bigl(U^{\prime}+4V^{\prime\prime}\bigr)(n^{\delta}+n^a)+4V^{\prime}n^a
\label{57}, \\
\epsilon_b=&\epsilon_d=\mu^{\prime}+\mu_1-\bigl(U^{\prime}+4V^{\prime\prime}\bigr)n^a+4V^{\prime}(n^{\delta}+n^a)
\label{58},\\
C^{\prime}=&C+N\biggl[U^{\prime}n^an^b-2V^{\prime}\Bigl\{{(n^a)}^2+{(n^b)}^2\Bigr\}+4V^{\prime\prime}n^an^b-8tI_0\bigl(\Delta-\Delta^{\dagger}\bigr)\bigl(\Delta^{\prime}- \Delta^{\prime\dagger}\bigr)\notag \\
&\qquad\quad+2V^{\prime\prime}\bigl(|\Delta|^2+|\Delta^{\prime}|^2\bigr)\biggr]
\label{59}
\end{align}
with $\mu_1=\mu_2.$

From Eq.\;(\ref{52}), we derive self-consistent equations for $\Delta$ and $\Delta^{\prime}$:
\begin{align}
\Delta=&\dfrac{1}{2N}\sum_{\bm{k}} \dfrac{s_{\bm{k}}^2\bigl\{V^{\prime\prime}\Delta+4tI_0(\Delta^{\prime}-\Delta^{\prime\dagger})\bigr\}}{\sqrt{(\epsilon_a+\epsilon_b)^2+4|\Gamma_{\bm{k}}|^2}}\bigl\{f(E_{\bm{k}}^{-})-f(E_{\bm{k}}^{ +})\bigr\}
\label{60},\\ 
\Delta^{\prime}=&\dfrac{1}{2N}\sum_{\bm{k}} \dfrac{s_{\bm{k}}^2\bigl\{V^{\prime\prime}\Delta^{\prime}+4t I_0(\Delta-\Delta^{\dagger})\bigl\}}{\sqrt{(\epsilon_a+\epsilon_b)^2+4|\Gamma_{\bm{k}}^{\prime}|^2}}\bigl\{f(E_{\bm{k}}^{\prime-})-f(E_{\bm{k}}^{\prime+})\bigl\}
\label{61},
\end{align}
where
\begin{align} 
E_{\bm{k}}^{\pm}=&\dfrac{1}{2}(\epsilon_a-\epsilon_b)\pm\dfrac{1}{2}\sqrt{(\epsilon_a+\epsilon_b)^2+4|\Gamma_{\bm{k}}|^2}
\label{62},\\
E_{\bm{k}}^{\prime\pm}=&-\dfrac{1}{2}(\epsilon_a-\epsilon_b)\pm\dfrac{1}{2}\sqrt{(\epsilon_a+\epsilon_b)^2+4|\Gamma_{\bm{k}}^{\prime}|^2} 
\label{63},
\end{align}
and $f(E)=\frac{1}{e^{\beta E}+1}$ is the Fermi distribution function. 
Furthermore, we self-consistently have $|\Gamma_{\bm{k}}^{\prime}|^2=|\Gamma_{\bm{k}}|^2$, $E_{\bm{k}}^{\prime\pm}=-E_{\bm{k}}^{\mp}$ and two types of solutions, which are given by: 
\begin{align}
\Delta=&\Delta^{\prime} \equiv i\Delta_{\rm{dw}}
\label{64}, \\
\Delta=&-\Delta^{\prime}\equiv i\tilde{\Delta}_{\rm{dw}}
\label{65}.
\end{align} 
In Appendix C, we describe in detail the derivation of the self-consistent equations.

In the case of $\Delta=\Delta^{\prime}$, the DDW of electrons with up spin is in phase with that of electrons with down spin. Therefore, we can detect the DDW.
From Eqs.\;(\ref{60}) or (\ref{61}), and using Eq.\;(\ref{53}), we obtain the magnitude of the DDW as:
 \begin{align}
1=\dfrac{v_{\rm{dw}}}{2N}&\sum_{\bm{k}}\dfrac{s_{\bm{k}}^2}{\sqrt{(\epsilon_a+\epsilon_b)^2+4\left\{t^2(1-m)^2\gamma_{\bm{k}}^2+v_{\rm{dw}}^2\Delta_{\rm{dw}}^2s_{\bm{k}}^2\right\}}}\left\{f(E_{{\bm{k}},1}^-)-f(E_{{\bm{k}},1}^+)\right\}
\label{66},
\end{align}
where
\begin{align}
v_{\rm{dw}}=&V^{\prime\prime}+8tI_0
\label{67},\\
E_{{\bm{k}},1}^{\pm}=&\dfrac{1}{2}(\epsilon_a-\epsilon_b)\pm\dfrac{1}{2}\sqrt{(\epsilon_a+\epsilon_b)^2+4\left\{t^2(1-m)^2\gamma_{\bm{k}}^2+v_{\rm{dw}}^2\Delta_{\rm{dw}}^2s_{\bm{k}}^2\right\}}
\label{67-2},
\end{align}
in addition, Eq.\;(\ref{67-2}) is obtained by substituting Eqs.\;(\ref{53}) and (\ref{64}) into Eq.\;(\ref{62}). 
At least $v_{\rm{dw}}>0$ is required for Eq.\;(\ref{66}) to have a solution because $f(E_{{\bm{k}},1}^-)>f(E_{{\bm{k}},1}^+)$, and the integrand of Eq.\;(\ref{66}) is always positive. 
Using Eqs.\;(\ref{12}), (\ref{13}), and (\ref{31}), Eq.\;(\ref{67}) is rewritten as: 
\begin{align}
v_{\rm{dw}}=\dfrac{4t^2}{U}\left\{\dfrac{x}{4}\biggl(\dfrac{U}{t}\biggr)^2+\dfrac{2-7x}{(2x-1)(x-1)}\right\}
\label{67-3},
\end{align}
where $x=V/U$. 
In the AF region (i.e., $x<1/4$), $v_{\rm{dw}}$ is positive definite. 
The integrand of Eq.\;(\ref{66}) is also positive definite. Therefore, Eq.\;(\ref{66}) may have a nontrivial solution $\Delta_{\rm{dw}}\neq0$.

By contrast, in the case of $\Delta=-\Delta^{\prime}$, the DDW of electrons with up spin is out of phase with that of electrons with down spin. Therefore, we cannot detect this type of DDW even if it exists. The magnitude of this type of DDW is given by: 
\begin{align}
1=\dfrac{\tilde{v}_{\rm{dw}}}{2N}&\sum_{\bm{k}}\dfrac{s_{\bm{k}}^2}{\sqrt{(\epsilon_a+\epsilon_b)^2+4\left\{t^2(1-m)^2\gamma_{\bm{k}}^2+\tilde{v}_{\rm{dw}}^2\tilde{\Delta}_{\rm{dw}}^2s_{\bm{k}}^2\right\}}}\left\{f(E_{{\bm{k}},2}^-)-f(E_{{\bm{k}},2}^+)\right\}
\label{68},
\end{align}
where
\begin{align}
\tilde{v}_{\rm{dw}}=&V^{\prime\prime}-8tI_0
\label{68-2},\\
E_{{\bm{k}},2}^{\pm}=&\dfrac{1}{2}(\epsilon_a-\epsilon_b)\pm\dfrac{1}{2}\sqrt{(\epsilon_a+\epsilon_b)^2+4\left\{t^2(1-m)^2\gamma_{\bm{k}}^2+\tilde{v}_{\rm{dw}}^2\tilde{\Delta}_{\rm{dw}}^2s_{\bm{k}}^2\right\}}
\label{68-3},
\end{align}
and Eq.\;(\ref{68-3}) is obtained by substituting Eqs.\;(\ref{53}) and (\ref{65}) into Eq.\;(\ref{62}). 
Furthermore, Eq.\;(\ref{68-2}) is rewritten as:
\begin{align}
\tilde{v}_{\rm{dw}}=\dfrac{4t^2}{U}\left\{\dfrac{x}{4}\biggl(\dfrac{U}{t}\biggr)^2+\dfrac{x-2}{(2x-1)(x-1)}\right\}
\label{69}.
\end{align}
If $U/t<8.64$, we have $\tilde{v}_{\rm{dw}}<0$ for the region of $x<1/4$. 
Because $f(E_{{\bm{k}},2}^-)>f(E_{{\bm{k}},2}^+)$ and the integrand of Eq.\;(\ref{68}) is always positive, Eq.\;(\ref{68}) has no solutions and we have $\Delta=-\Delta^{\prime}=0$ when $U/t<8.64$. Fortunately, many substances of HTS satisfy this condition. Thus, we need not consider the case of $\Delta=-\Delta^{\prime}$. 
Therefore, in the following, we consider only the case of $\Delta=\Delta^{\prime}=i\Delta_{\rm{dw}}$. 

We must obtain the ground state energy to examine whether the solution of $\Delta=\Delta^{\prime}$ is stable. 
From Eq.\;(\ref{43}), we obtain: 
\begin{align}
\dfrac{\langle \mathcal{H}_{\rm{DDW}}^{\rm{MF}}\rangle}{N}=E(n^a,n^b)+E(\Delta_{\rm{dw}})
\label{70},
\end{align}
where
\begin{align}
E(n^a,n^b)=&\frac{C}{N}+\epsilon_an^a+\epsilon_bn^b+(U^{\prime}+4V^{\prime\prime})n^an^b-2V^{\prime}\left\{(n^a)^2+(n^b)^2\right\}
\label{71},\\
E(\Delta_{\rm{dw}})=&-4v_{\rm{dw}}\Delta_{\rm{dw}}^2
\label{72}.
\end{align}
We note that Eq.\;(\ref{71}) depends only on $n^{\delta}$ because $n^a=\frac{1-n^{\delta}}{2}$ and $n^b=\frac{1+n^{\delta}}{2}$. 
It is determined only by Eq.\;(\ref{72}) whether the DDW is stable. 
As previously mentioned, we obtained $v_{\rm{dw}}>0$ and $\Delta_{\rm{dw}}\neq0$. Therefore, we can conclude that the DDW is always stable.

We derive the equation for obtaining the DDW order parameter $\Delta_{\rm{dw}}$ in the nonmagnetic phase ($m=0$), as later it is proved that the magnetic phase is allowed only at $n^{\delta}=0$ in our approximation. 
From Eqs.\;(\ref{c12}) and (\ref{c20}), we obtain:
\begin{align}
n^b=&\frac{2}{N}\sum_{\bm{k}}\langle b_{\bm{k}}^{\dagger}b_{\bm{k}}\rangle \notag \\
=&1-\dfrac{2}{N}\sum_{\bm{k}}\dfrac{\epsilon_a+\epsilon_b}{2\sqrt{(\epsilon_a+\epsilon_b)^2+4\left\{t^2(1-m)^2\gamma_{\bm{k}}^2+v_{\rm{dw}}^2\Delta_{\rm{dw}}^2s_{\bm{k}}^2\right\}}}\left\{f(E_{{\bm{k}},1}^-)-f(E_{{\bm{k}},1}^+)\right\}\notag \\
&\;-\dfrac{2}{N}\sum_{\bm{k}}\dfrac{1}{2}\left\{f(E_{{\bm{k}},1}^+)+f(E_{{\bm{k}},1}^-)\right\}
\label{73},\\ 
n^a=&\frac{2}{N}\sum_{\bm{k}}\langle a_{\bm{k}}^{\dagger}a_{\bm{k}}\rangle \notag \\
=&-\dfrac{2}{N}\sum_{\bm{k}}\dfrac{\epsilon_a+\epsilon_b}{2\sqrt{(\epsilon_a+\epsilon_b)^2+4\left\{t^2(1-m)^2\gamma_{\bm{k}}^2+v_{\rm{dw}}^2\Delta_{\rm{dw}}^2s_{\bm{k}}^2\right\}}}\left\{f(E_{{\bm{k}},1}^-)-f(E_{{\bm{k}},1}^+)\right\}\notag \\
&\;+\dfrac{2}{N}\sum_{\bm{k}}\dfrac{1}{2}\left\{f(E_{{\bm{k}},1}^+)+f(E_{{\bm{k}},1}^-)\right\}
\label{74},
\end{align}
where we have utilized Eqs.\;(\ref{53}), (\ref{62}), (\ref{64}), and (\ref{c35}). 
Therefore, from Eqs.\;(\ref{73}) and (\ref{74}), the hole concentration $n^{\delta}$ is obtained by:
\begin{align}
n^{\delta}=&n^b-n^a\notag \\
	   =&1-\dfrac{2}{N}\sum_{\bm{k}} \left\{f(E_{{\bm{k}},1}^+)+f(E_{{\bm{k}},1}^-)\right\}
\label{75}.
\end{align}
From Eqs.\;(\ref{57}) and (\ref{58}), we derive:
\begin{align}
\epsilon_a+\epsilon_b=&\left\{U+16t(2I_0-3I_1-3I_2)\right\}(1-n^{\delta}-2n^a)\notag \\
=&v_{m}m
\label{76},
\end{align}
where $v_{m}=U+16t(2I_0-3I_1-3I_2)$, and we used Eqs.\;(\ref{27})--(\ref{31}) and (\ref{34}). 
Furthermore, from Eqs.\;(\ref{74})--(\ref{76}), the magnetization $m$ is self-consistently given by: 
\begin{align}
m=&1-n^{\delta}-2n^a\notag \\
=&\dfrac{2}{N}\sum_{\bm{k}}\dfrac{\epsilon_a+\epsilon_b}{\sqrt{(\epsilon_a+\epsilon_b)^2+4\left\{t^2(1-m)^2\gamma_{\bm{k}}^2+v_{\rm{dw}}^2\Delta_{\rm{dw}}^2s_{\bm{k}}^2\right\}}}\left\{f(E_{{\bm{k}},1}^-)-f(E_{{\bm{k}},1}^+)\right\} \notag \\
=&v_{m} m\dfrac{2}{N}\sum_{\bm{k}}\dfrac{f(E_{{\bm{k}},1}^-)-f(E_{{\bm{k}},1}^+)}{\sqrt{(v_{m}m)^2+4\left\{t^2(1-m)^2\gamma_{\bm{k}}^2+v_{\rm{dw}}^2\Delta_{\rm{dw}}^2s_{\bm{k}}^2\right\}}}
\label{77},
\end{align}
and the Neel temperature $T_N$, which is defined by $m=0$, is given by:
\begin{eqnarray}
1=v_{m}\dfrac{2}{N}\sum_{\bm{k}}\dfrac{\left\{f(E_{{\bm{k}},1,m=0}^-)-f(E_{{\bm{k}},1,m=0}^+)\right\}|_{T=T_N}}{2\sqrt{t^2\gamma_{\bm{k}}^2+v_{\rm{dw}}^2\Delta_{\rm{dw}}^2s_{\bm{k}}^2}}
\label{78},
\end{eqnarray}
where 
\begin{eqnarray}
E_{{\bm{k}},1,m=0}^{\pm}=&\epsilon_a\pm\sqrt{t^2\gamma_{\bm{k}}^2+v_{\rm{dw}}^2\Delta_{\rm{dw}}^2s_{\bm{k}}^2}
\label{78-2},
\end{eqnarray}
and Eq.\;(\ref{78-2}) is obtained by substituting $\epsilon_a+\epsilon_b=0$ and $m=0$ into Eq.\;(\ref{67-2}). 
Note that, from Eq.\;(\ref{76}), we have $\epsilon_a+\epsilon_b=0$ when $m=0$. 

Substituting $m=0$ and Eq.\;(\ref{78-2}) into Eq.\;(\ref{66}), the DDW order parameter $\Delta_{\rm{dw}}$ in the nonmagnetic phase is given by: 
\begin{eqnarray}
1=\dfrac{v_{\rm{dw}}}{4N}\sum_{\bm{k}}\dfrac{s_{\bm{k}}^2}{\sqrt{t^2\gamma_{\bm{k}}^2+v_{\rm{dw}}^2\Delta_{\rm{dw}}^2s_{\bm{k}}^2}}\left\{f(E_{{\bm{k}},1,m=0}^-)-f(E_{{\bm{k}},1,m=0}^+)\right\}
\label{79}.
\end{eqnarray}

Finally, we conduct the diagonalization of Eq.\;(\ref{52}) and write as: 
\begin{align}
\Gamma_{\bm{k}}=\Gamma_{\bm{k}}^0 e^{-i\lambda_{\bm{k}}}
\label{80}, 
\end{align}
where
\begin{align}
&\Gamma_{\bm{k}}^0=\sqrt{t^2(1-m)^2\gamma_{\bm{k}}^2+v_{\rm{dw}}^2\Delta_{\rm{dw}}^2s_{\bm{k}}^2}
\label{81},\\
&\tan(\lambda_{\bm{k}})=\dfrac{v_{\rm{dw}}\Delta_{\rm{dw}}s_{\bm{k}}}{t(1-m)\gamma_{\bm{k}}}
\label{82}.
\end{align}
Substituting Eq.\;(\ref{80}) into Eq.\;(\ref{52}) and replacing 
\begin{align}
a_{\bm{k}}\rightarrow e^{-i\lambda_{\bm{k}}/2}a_{\bm{k}},\;d_{\bm{k}}\rightarrow e^{-i\lambda_{\bm{k}}/2}d_{\bm{k}},\;b_{\bm{k}} \rightarrow e^{i\lambda_{\bm{k}}/2}b_{\bm{k}},\;c_{\bm{k}} \rightarrow e^{i\lambda_{\bm{k}}/2}c_{\bm{k}}
\label{83},
\end{align}
we obtain
\begin{align}
\mathcal{H}_{\rm{DDW}}^{\rm{MF}}=&\sum_{\bm{k}} \Bigl\{\epsilon_a\bigl(a_{\bm{k}}^{\dagger}a_{\bm{k}}+c_{\bm{k}}^{\dagger}c_{\bm{k}}\bigr)+\epsilon_b\bigl(b_{\bm{k}}^{\dagger}b_{\bm{k}}+d_{\bm{k}}^{\dagger}d_{\bm{k}}\bigr)
+\Gamma_{\bm{k}}^0\bigl(a_{\bm{k}}d_{\bm{k}}+d_{\bm{k}}^{\dagger}a_{\bm{k}}^{\dagger}-b_{\bm{k}}c_{\bm{k}}-c_{\bm{k}}^{\dagger}b_{\bm{k}}^{\dagger}\bigr)\Bigr\},
\label{84}
\end{align}
and omit the constant term.
Furthermore, the canonical transformation is applied:
\begin{align}
a_{\bm{k}}=&\cos \theta_{\bm{k}} A_{\bm{k}}-\sin \theta_{\bm{k}} D_{\bm{k}}^{\dagger}
\label{85}, \\
d_{\bm{k}}^{\dagger}=&\sin \theta_{\bm{k}} A_{\bm{k}}+\cos \theta_{\bm{k}} D_{\bm{k}}^{\dagger}
\label{86}, \\
c_{\bm{k}}=&\cos \theta_{\bm{k}} C_{\bm{k}}- \sin \theta_{\bm{k}} B_{\bm{k}}^{\dagger}
\label{87}, \\
b_{\bm{k}}^{\dagger}=&\sin \theta_{\bm{k}} C_{\bm{k}}+\cos \theta_{\bm{k}} B_{\bm{k}}^{\dagger}
\label{88},
\end{align}
where
\begin{align}
\cos 2\theta_{\bm{k}}=&\dfrac{\epsilon_a+\epsilon_b}{\sqrt{(\epsilon_a+\epsilon_b)^2+4(\Gamma_{\bm{k}}^0})^2}
\label{89},\\
\sin 2 \theta_{\bm{k}}=&\dfrac{-2(\Gamma_{\bm{k}}^0)^2}{\sqrt{(\epsilon_a+\epsilon_b)^2+4(\Gamma_{\bm{k}}^0)^2}}
\label{90}.
\end{align}
Substituting these transformation into  Eq.\;(\ref{84}), we finally obtain the diagonal mean-field Hamiltonian:
\begin{align}
\mathcal{H}_{\rm{DDW}}^{\rm{MF}}=&\sum_{\bm{k}}\dfrac{1}{2}\left\{\sqrt{(\epsilon_a+\epsilon_b)^2+4(\Gamma_{\bm{k}}^0)^2}+(\epsilon_a-\epsilon_b)\right\}
\bigl(A_{\bm{k}}^{\dagger} A_{\bm{k}}+C_{\bm{k}}^{\dagger}C_{\bm{k}}\bigr) \notag \\
&+\sum_{\bm{k}}\dfrac{1}{2}\left\{\sqrt{(\epsilon_a+\epsilon_b)^2+4(\Gamma_{\bm{k}}^0)^2}-(\epsilon_a-\epsilon_b)\right\}
\bigl(D_{\bm{k}}^{\dagger}D_{\bm{k}}+B_{\bm{k}}^{\dagger}B_{\bm{k}}\bigr) \notag \\
&+\sum_{\bm{k}}\left\{(\epsilon_a+\epsilon_b)-\sqrt{(\epsilon_a+\epsilon_b)^2+4(\Gamma_{\bm{k}}^0)^2} \right\}
\label{91}.
\end{align}
In the nonmagnetic phase, we set $m=0$ and obtain:
\begin{align}
\mathcal{H}_{\rm{DDW},\text{$m=0$}}^{\rm{MF}}=&\sum_{\bm{k}}\left(\sqrt{t^2\gamma_{\bm{k}}^2+v_{\rm{dw}}^2\Delta_{\rm{dw}}^2s_{\bm{k}}^2}+\epsilon_a\right)
\bigl(A_{\bm{k}}^{\dagger}A_{\bm{k}}+C_{\bm{k}}^{\dagger}C_{\bm{k}}\bigr) \notag \\
&+\sum_{\bm{k}} \left(\sqrt{t^2\gamma_{\bm{k}}^2+v_{\rm{dw}}^2\Delta_{\rm{dw}}^2s_{\bm{k}}^2}-\epsilon_a\right)\bigl(D_{\bm{k}}^{\dagger}D_{\bm{k}}+B_{\bm{k}}^{\dagger}B_{\bm{k}}\bigr)-2\sum_{\bm{k}}\sqrt{t^2\gamma_{\bm{k}}^2+v_{\rm{dw}}^2\Delta_{\rm{dw}}^2s_{\bm{k}}^2}
\label{92},
\end{align}
where
\begin{equation}
\epsilon_a=\dfrac{1}{2}(U+8V)-\dfrac{n^{\delta}}{2}\left\{U+8V-48t(I_1-I_2)\right\}-\mu_1
\label{93}.
\end{equation}

\section{$d_{x^2-y^2}$-wave superconducting phase}

The superconducting dome of HTS is located in the pseudogap phase, which is considered the DDW phase in the phase diagram.\cite{ref.Chakravarty} 
Therefore, we consider the superconductivity with the DDW order parameters. 
Furthermore, the experiments show that the Cooper pair is the bound state of holes and the superconductivity is a $d_{x^2-y^2}$ type in the nonmagnetic phase.\cite{ref.experiment,ref.ZXShen,ref.H.Ding} 
Therefore, we consider the DSC phase, which is established based on the DDW order parameters. 
First, we introduce the DSC order parameters: 
\begin{eqnarray}
\langle d_{i+\delta}b_i \rangle=\begin{cases}
\langle d_{i\pm \bm{e}_x}b_i \rangle = \Delta_{\rm{ds}} \\
 \langle d_{i\pm \bm{e}_y} b_i \rangle =- \Delta_{\rm{ds}} \label{377}
\end{cases}
\label{94},
\end{eqnarray}
and 
\begin{eqnarray}
\langle c_{i+\delta}^{\dagger} a_i^{\dagger} \rangle =\begin{cases}
\langle c_{i\pm \bm{e}_x}^{\dagger}a_i^{\dagger} \rangle = \Delta_{\rm{ds}}^{\prime} \\
\langle c_{i\pm \bm{e}_y}^{\dagger}a_i^{\dagger} \rangle =- \Delta_{\rm{ds}}^{\prime}
\end{cases}
\label{95}.
\end{eqnarray}

Next, we obtain the DSC part of the Hamiltonian. 
Performing the mean-field approximation in the Hamiltonian Eq.\;(\ref{21}) by using Eqs.\;(\ref{94}) and (\ref{95}), the mean-field Hamiltonian of the DSC part is written as:
\begin{align}
\mathcal{H}_{\Delta_{\rm{ds}}}^{\rm{MF}}=&\sum_i\Bigl\{\bigl(V^{\prime}\Delta_{\rm{ds}}+4tI_0\Delta_{\rm{ds}}^{\prime}\bigr)\bigl(b_i^{\dagger}d_{i+\bm{e}_x}^{\dagger}+b_i^{\dagger}d_{i-\bm{e}_x}^{\dagger}-b_i^{\dagger}d_{i+\bm{e}_y}^{\dagger}-b_i^{\dagger}d_{i-\bm{e}_y}^{\dagger}\bigr)\notag \\
&\quad\;+\bigl(V^{\prime}\Delta_{\rm{ds}}^{\dagger}+4tI_0\Delta_{\rm{ds}}^{\prime\dagger}\bigr)\bigl(d_{i+\bm{e}_x}b_{i}+d_{i-\bm{e}_x}b_{i}-d_{i+\bm{e}_y}b_{i}-d_{i-\bm{e}_y}b_{i}\bigr)\notag \\
&\quad\;+\bigl(V^{\prime}\Delta_{\rm{ds}}^{\prime\dagger}+4tI_0\Delta_{\rm{ds}}^{\dagger}\bigr)\bigl(c_{i+\bm{e}_x}^{\dagger}a_{i}^{\dagger}+c_{i-\bm{e}_x}^{\dagger}a_{i}^{\dagger}-c_{i+\bm{e}_y}^{\dagger}a_{i}^{\dagger}-c_{i-\bm{e}_y}^{\dagger}a_{i}^{\dagger}\bigr)\notag \\
&\quad\;+\bigl(V^{\prime}\Delta_{\rm{ds}}^{\prime}+4tI_0\Delta_{\rm{ds}}\bigr)\bigl(a_ic_{i+\bm{e}_x}+a_ic_{i-\bm{e}_x}-a_ic_{i+\bm{e}_y}-a_ic_{i-\bm{e}_y}\bigr)\Bigr\}\notag \\
&\quad\;-2N\Bigl\{V^{\prime}\bigl(|\Delta_{\rm{ds}}|^2+|\Delta_{\rm{ds}}^{\prime}|^2\bigr)+4tI_0\bigl(\Delta_{\rm{ds}}^{\prime\dagger}\Delta_{\rm{ds}}+\Delta_{\rm{ds}}^{\dagger}\Delta_{\rm{ds}}^{\prime}\bigr)\Bigr\}
\label{96-0}.
\end{align}
Furthermore, performing the Fourier transformation, we can rewrite Eq.\;(\ref{96-0}) as:
\begin{align}
\mathcal{H}_{\Delta_{\rm{ds}}}^{\rm{MF}}=\sum_{\bm{k}} s_{\bm{k}}\bigl(\Lambda b_{\bm{k}}^{\dagger}d_{\bm{k}}^{\dagger}+\Lambda^{\dagger}d_{\bm{k}}b_{\bm{k}}+\Lambda^{\prime}c_{\bm{k}}^{\dagger}a_{\bm{k}}^{\dagger}+\Lambda^{\prime\dagger}a_{\bm{k}} c_{\bm{k}}\bigr) +C^{\prime\prime}
\label{96},
\end{align}
where
\begin{align}
\Lambda=&V^{\prime}\Delta_{\rm{ds}}+4t I_0 \Delta_{\rm{ds}}^{\prime}
\label{97},\\
\Lambda^{\prime}=&V^{\prime}\Delta_{\rm{ds}}^{\prime\dagger}+4t I_0 \Delta_{\rm{ds}}^{\dagger}
\label{98},\\
C^{\prime\prime}=&-2N\Bigl\{V^{\prime}\bigl(|\Delta_{\rm{ds}}|^2+|\Delta_{\rm{ds}}^{\prime}|^2\bigr)+4tI_0\bigl(\Delta_{\rm{ds}}^{\prime\dagger}\Delta_{\rm{ds}}+\Delta_{\rm{ds}}^{\dagger}\Delta_{\rm{ds}}^{\prime}\bigr)\Bigr\}
\label{99}.
\end{align}
Therefore, we obtain the total Hamiltonian by adding the previous Hamiltonian  Eq.\;(\ref{52}) to Hamiltonian  Eq.\;(\ref{96}):
\begin{align}
\mathcal{H}_{\rm{tot}}^{\rm{MF}}=&\mathcal{H}_{\rm{DDW}}^{\rm{MF}}+\mathcal{H}_{\Delta_{\rm{ds}}}^{\rm{MF}}\notag \\
=&\sum_{\bm{k}}\biggl\{\epsilon_a\bigl(a_{\bm{k}}^{\dagger}a_{\bm{k}}+c_{\bm{k}}^{\dagger}c_{\bm{k}}\bigr)+\epsilon_b\bigl(b_{\bm{k}}^{\dagger}b_{\bm{k}}+d_{\bm{k}}^{\dagger}d_{\bm{k}}\bigr)\notag \\
&\qquad+\sqrt{t^2(1-m)^2\gamma_{\bm{k}}^2+v_{\rm{dw}}^2\Delta_{\rm{dw}}^2s_{\bm{k}}^2}\bigl(a_{\bm{k}} d_{\bm{k}}+d_{\bm{k}}^{\dagger}a_{\bm{k}}^{\dagger}-b_{\bm{k}}c_{\bm{k}}-c_{\bm{k}}^{\dagger}b_{\bm{k}}^{\dagger}\bigr) \notag \\
&\qquad+s_{\bm{k}}\bigl(\Lambda b_{\bm{k}}^{\dagger}d_{\bm{k}}^{\dagger}+\Lambda^{\dagger}d_{\bm{k}} b_{\bm{k}}+\Lambda^{\prime}c_{\bm{k}}^{\dagger}a_{\bm{k}}^{\dagger}+\Lambda^{\prime\dagger}a_{\bm{k}}c_{\bm{k}}\bigr)\biggr\} +C^{\prime}+C^{\prime\prime}
\label{100},
\end{align}
where we have utilized Eqs.\;(\ref{80}), (\ref{81}), and (\ref{83}).
The experimental results suggest that superconductivity is established on the nonmagnetic phase.\cite{ref.experiment} 
Therefore, $\mathcal{H}_{\rm{tot}}^{\rm{MF}}$ with $m=0$ is the Hamiltonian of the DSC phase, which coexists with the DDW phase. 
Therefore, substituting $m=0$ ($\epsilon_a+\epsilon_b=0$) into Eqs.\;(\ref{89}) and (\ref{90}), $\theta_{\bm{k}}=-\pi/4$ is obtained and Eqs.\;(\ref{85})--(\ref{88}) are written as:
\begin{align}
a_{\bm{k}}=&\dfrac{1}{\sqrt{2}}\bigl(A_{\bm{k}}+D_{\bm{k}}^{\dagger}\bigr)
\label{101},\\
d_{\bm{k}}^{\dagger}=&\dfrac{1}{\sqrt{2}}\bigl(-A_{\bm{k}}+ D_{\bm{k}}^{\dagger}\bigr)
\label{102},\\
c_{\bm{k}}=&\dfrac{1}{\sqrt{2}}\bigl(C_{\bm{k}}+ B_{\bm{k}}^{\dagger}\bigr)
\label{103},\\
b_{\bm{k}}^{\dagger}=&\dfrac{1}{\sqrt{2}}\bigl(-C_{\bm{k}}+B_{\bm{k}}^{\dagger}\bigr)
\label{104}.
\end{align}
Using Eqs.\;(\ref{94}),(\ref{95}), and (\ref{101})--(\ref{104}), we obtain:
\begin{align}
\Delta_{\rm{ds}}=&\dfrac{1}{2N}\sum_{\bm{k}} s_{\bm{k}}\langle d_{\bm{k}}b_{\bm{k}}\rangle\notag \\
	    =&\dfrac{1}{2N} \sum_{\bm{k}} s_{\bm{k}}\dfrac{1}{2}\bigl(\langle A_{\bm{k}}^{\dagger}C_{\bm{k}}^{\dagger}\rangle+\langle D_{\bm{k}}B_{\bm{k}}\rangle+\langle C_{\bm{k}}^{\dagger}D_{\bm{k}}\rangle-\langle A_{\bm{k}}^{\dagger}B_{\bm{k}} \rangle\bigr)
	   \label{105},\\
\Delta_{\rm{ds}}^{\prime}=&\dfrac{1}{2N}\sum_{\bm{k}} s_{\bm{k}}\langle c_{\bm{k}}^{\dagger}a_{\bm{k}}^{\dagger}\rangle\notag \\
	    =&-\dfrac{1}{2N} \sum_{\bm{k}} s_{\bm{k}}\dfrac{1}{2}\bigl(\langle A_{\bm{k}}^{\dagger}C_{\bm{k}}^{\dagger}\rangle+\langle D_{\bm{k}}B_{\bm{k}}\rangle-\langle C_{\bm{k}}^{\dagger}D_{\bm{k}}\rangle+\langle A_{\bm{k}}^{\dagger}B_{\bm{k}} \rangle\bigr)
	   \label{106},
\end{align}
and assume that two types of solutions are possible. One of them is given by $\Delta_{\rm{ds}}=\Delta_{\rm{ds}}^{\prime}$ and the other is $ \Delta_{\rm{ds}}=-\Delta_{\rm{ds}}^{\prime}$. 
When $\Delta_{\rm{ds}}=\Delta_{\rm{ds}}^{\prime}$, Appendix D shows that this case can be permitted only in the region with $4V<U<\dfrac{1}{2}(V+\sqrt{V^2+48t^2})$, that is, roughly speaking, $0<V/t<1$ and $U/t<4$. 
Therefore, when $4<U/t$, considering this solution is not necessary. 

As discussed in the following, we consider only the $\Delta_{\rm{ds}}=-\Delta_{\rm{ds}}^{\prime}$ case. 
Therefore, from Eqs.\;(\ref{97}) and (\ref{98}), we obtain:
\begin{align}
\Lambda=-\Lambda^{\prime\dagger}=(V^{\prime}-4tI_0)\Delta_{\rm{ds}}
\label{107}.
\end{align}
Substituting $m=0$ and Eqs.\;(\ref{101})--(\ref{104}) and (\ref{107}) into  Eq.\;(\ref{100}), then, $\mathcal{H}_{\rm{tot}}^{\rm{MF}}$ with $m=0$ can be considered as the Hamiltonian of the DSC phase, which is written as:
\begin{align}
\mathcal{H}_{\rm{DSC}}^{\rm{MF}}=&\sum_{\bm{k}}\Bigl\{\mathcal{E}_{\bm{k}}^+\bigl(A_{\bm{k}}^{\dagger}A_{\bm{k}}+C_{\bm{k}}^{\dagger}C_{\bm{k}}\bigr)+\mathcal{E}_{\bm{k}}^-\bigl(B_{\bm{k}}^{\dagger}B_{\bm{k}}+D_{\bm{k}}^{\dagger}D_{\bm{k}}\bigr) \notag \\
&\quad+s_{\bm{k}}\Lambda\bigl(C_{\bm{k}}A_{\bm{k}}+B_{\bm{k}}^{\dagger}D_{\bm{k}}^{\dagger}+A_{\bm{k}}^{\dagger}C_{\bm{k}}^{\dagger}+D_{\bm{k}}B_{\bm{k}}\bigr)\Bigr\}
\label{108},
\end{align}
where 
\begin{align}
\mathcal{E}_{\bm{k}}^{\pm}=&\sqrt{t^2\gamma_{\bm{k}}^2+v_{\rm{dw}}^2\Delta_{\rm{dw}}^2s_{\bm{k}}^2}\pm \epsilon_a
\label{109}.
\end{align}
Here, we have treated $\Lambda$ as a real number because its phase can be absorbed into operators, and have omitted the constant term. 
Furthermore, the canonical transformation is applied:
\begin{eqnarray}
A_{\bm{k}}=&\cos\phi_{\bm{k}}^{+}\alpha_{\bm{k}}^{+}-\sin \phi_{\bm{k}}^{+} \beta_{\bm{k}}^{+\dagger}
\label{110},\\
C_{\bm{k}}^{\dagger}=&\sin\phi_{\bm{k}}^+\alpha_{\bm{k}}^++\cos \phi_{\bm{k}}^+ \beta_{\bm{k}}^{+\dagger}
\label{111},\\
B_{\bm{k}}=&\cos\phi_{\bm{k}}^-\alpha_{\bm{k}}^--\sin\phi_{\bm{k}}^- \beta_{\bm{k}}^{-\dagger}
\label{112},\\
D_{\bm{k}}^{\dagger}=&\sin\phi_{\bm{k}}^-\alpha_{\bm{k}}^-+\cos\phi_{\bm{k}}^-\beta_{\bm{k}}^{-\dagger}
\label{113},
\end{eqnarray}
where
\begin{align}
\cos 2\phi_{\bm{k}}^{\pm} =\dfrac{\mathcal{E}_{\bm{k}}^{\pm}}{\sqrt{(\mathcal{E}_{\bm{k}}^{\pm})^2+s_{\bm{k}}^2\Lambda^2}}
\label{114},\\
\sin 2\phi_{\bm{k}}^{\pm}=\dfrac{s_{\bm{k}}\Lambda}{\sqrt{(\mathcal{E}_{\bm{k}}^{\pm})^2+s_{\bm{k}}^2\Lambda^2}}
\label{115}.
\end{align}
Substituting these transformation into  Eq.\;(\ref{108}), we finally obtain the diagonal mean-field Hamiltonian:
\begin{align}
\mathcal{H}_{\rm{DSC}}^{\rm{MF}}=&\sum_{\bm{k}}\biggl[\sqrt{\bigl(\mathcal{E}_{\bm{k}}^{-}\bigr)^2+s_{\bm{k}}^2\Lambda^2}\bigl(\alpha_{\bm{k}}^{-\dagger}\alpha_{\bm{k}}^{-}+\beta_{\bm{k}}^{-\dagger}\beta_{\bm{k}}^{-}\bigr)+\Bigl\{\mathcal{E}_{\bm{k}}^{-}-\sqrt{(\mathcal{E}_{\bm{k}}^{-})^2+s_{\bm{k}}^2\Lambda^2} \Bigr\} \notag \\
&\qquad+\sqrt{\bigl(\mathcal{E}_{\bm{k}}^{+}\bigr)^2+s_{\bm{k}}^2\Lambda^2}\bigl(\alpha_{\bm{k}}^{+\dagger}\alpha_{\bm{k}}^{+}+\beta_{\bm{k}}^{+\dagger}\beta_{\bm{k}}^{+}\bigr)+\Bigl\{\mathcal{E}_{\bm{k}}^{+}-\sqrt{(\mathcal{E}_{\bm{k}}^{+})^2+ s_{\bm{k}}^2\Lambda^2}\Bigr\}\biggr]
\label{116}.
\end{align}
Note that, from Eqs.\;(\ref{105}) and (\ref{106}), our assumption $\Delta_{\rm{ds}}=-\Delta_{\rm{ds}}^{\prime}$ is equivalent to the relation $\langle C_{\bm{k}}^{\dagger}D_{\bm{k}}\rangle=\langle A_{\bm{k}}^{\dagger}B_{\bm{k}}\rangle$. 
However, from Eqs.\;(\ref{110})--(\ref{113}) and (\ref{116}), we find that $\langle C_{\bm{k}}^{\dagger}D_{\bm{k}}\rangle=\langle A_{\bm{k}}^{\dagger}B_{\bm{k}}\rangle=0$, which satisfies the aforementioned relation. This result means that $\Delta_{\rm{ds}}=-\Delta_{\rm{ds}}^{\prime}$ is self-consistently satisfied. 

Substituting Eq.\;(\ref{110})--(\ref{113}) into Eq.\;(\ref{105}), the self-consistent equation for $\Delta_{\rm{ds}}$ is given as: 
\begin{align}
\Delta_{\rm{ds}}=&-\dfrac{v_{\rm{ds}}\Delta_{\rm{ds}}}{8N}\sum_{\bm{k}}\Biggl\{\dfrac{s_{\bm{k}}^2}{\sqrt{(\mathcal{E}_{\bm{k}}^{-})^2+v_{\rm{ds}}^2\Delta_{\rm{ds}}^2 s_{\bm{k}}^2}}\tanh\dfrac{\beta \sqrt{(\mathcal{E}_{\bm{k}}^{-})^2+v_{\rm{ds}}^2\Delta_{\rm{ds}}^2s_{\bm{k}}^2}}{2} \notag \\ 
&\qquad\qquad\qquad+\dfrac{s_{\bm{k}}^2}{\sqrt{(\mathcal{E}_{\bm{k}}^{+})^2+v_{\rm{ds}}^2\Delta_{\rm{ds}}^2s_{\bm{k}}^2}}\tanh\dfrac{\beta\sqrt{(\mathcal{E}_{\bm{k}}^{+})^2+v_{\rm{ds}}^2\Delta_{\rm{ds}}^2s_{\bm{k}}^2}}{2}\Biggr\}
\label{117},
\end{align}
where
\begin{align}
v_{\rm{ds}}=V^{\prime}-4tI_0=V-4t^2\left(\frac{5}{U-V}-\frac{3}{U}\right)
\label{118}.
\end{align}
Because the integrand of Eq.\;(\ref{117}) is always positive, the condition for Eq.\;(\ref{117}) to have a solution is $v_{\rm{ds}}<0$, which is written as $t^2>\frac{(2U+3V)V}{U(U-V)}$. 
This means that $t^2$ mainly contributes to the emergence of the DSC. 
Furthermore, using Eqs.\;(\ref{12}), (\ref{14}), and (\ref{30}), we can rewrite Eq.\;(\ref{118}) as:
\begin{align}
v_{\rm{ds}}=\dfrac{4t^2}{U}\left\{\dfrac{x}{4}\biggl(\dfrac{U}{t}\biggr)^2+\dfrac{3x+2}{x-1}\right\}
\label{119}.
\end{align}

\section{Results and Discussion}

In this section, we derive the phase diagrams by solving the self-consistent equations numerically. 
Furthermore, we obtain the Fermi surface, the density of states, and discuss the physical properties in the DDW state. 

\subsection{Phase diagrams}

We numerically solve the self-consistent Eqs.\;(\ref{66}) and (\ref{117}) to obtain the DDW and DSC order parameters ($\Delta_{\rm{dw}}$ and $\Delta_{\rm{ds}}$) in the nonmagnetic phase, and herein describe the phase diagrams. 
For convenience sake, we can write the concrete form of the self-consistent equations as follows.

From Eqs.\;(\ref{66}) and (\ref{76}), $\Delta_{\rm{dw}}$ are obtained by:
\begin{align}
1=\dfrac{v_{\rm{dw}}}{16\pi^2}\int_{-\pi}^{\pi}\int_{-\pi}^{\pi}dk_xdk_y\dfrac{(\cos k_x-\cos k_y)^2\left\{f(E_{\bm{k},1}^{-})-f(E_{\bm{k},1}^{+})\right\}}{\sqrt{(v_{m}m/4)^2+t^2(1-m)^2(\cos k_x+\cos k_y)^2+v_{\rm{dw}}^2\Delta_{\rm{dw}}^2(\cos k_x-\cos k_y)^2}}
\label{120},
\end{align}
where
\begin{align}
f(E_{\bm{k},1}^{\pm})=\frac{1}{\exp\Bigl\{2\beta\bigl(\epsilon_{\rm{F}}\pm \sqrt{(v_{m}m/4)^2+t^2(1-m)^2(\cos k_x+\cos k_y)^2+v_{\rm{dw}}^2\Delta_{\rm{dw}}^2(\cos k_x-\cos k_y)^2\bigr)}\Bigr\} +1}
\label{121}.
\end{align}
The integration is carried out over the first Brillouin zone of the square lattice, and $\epsilon_{\rm{F}}=(\epsilon_a-\epsilon_b)/4$ is determined to be the effective chemical potential. 
Furthermore, from Eq.\;(\ref{77}), the magnetizations are obtained by 
\begin{align}
1=\dfrac{v_{m}}{16\pi^2}\int_{-\pi}^{\pi}\int_{-\pi}^{\pi}dk_xdk_y\dfrac{f(E_{\bm{k},1}^{-})-f(E_{\bm{k},1}^{+})}{\sqrt{(v_{m}m/4)^2+t^2(1-m)^2(\cos k_x+\cos k_y)^2+v_{\rm{dw}}^2\Delta_{\rm{dw}}^2(\cos k_x-\cos k_y)^2}}
\label{122},
\end{align}
and from (\ref{75}), the hole concentration $n^{\delta}$ is written as:
\begin{align} 
n^{\delta}=1-\dfrac{1}{4\pi^2}\int_{-\pi}^{\pi}\int_{-\pi}^{\pi}dk_xdk_y\left\{f(E_{\bm{k},1}^{-})+f(E_{\bm{k},1}^{+})\right\}
\label{123}.
\end{align}

First, we consider when $n^{\delta}=0$ at a zero temperature, where the AF ground state is expected. 
In this case, we have $f(E_{\bm{k},1}^{-})=1$, $f(E_{\bm{k},1}^{+})=0$ for all $\bm{k}$, and $\epsilon_{\rm{F}}=0$ because of $E_{\bm{k},1}^{-}<E_{\bm{k},1}^{+}$. Therefore, Eq.\;(\ref{122}) is rewritten as:
\begin{align}
1=\dfrac{v_{m}}{16\pi^2}\int_{-\pi}^{\pi}\int_{-\pi}^{\pi}dk_xdk_y\dfrac{1}{\sqrt{(v_{m}m/4)^2+t^2(1-m)^2(\cos k_x+\cos k_y)^2+v_{\rm{dw}}^2\Delta_{\rm{dw}}^2(\cos k_x-\cos k_y)^2}}
\label{124}
\end{align}
and we notice that Eq.\;(\ref{124}) is satisfied only in the case of $\Delta_{\rm{dw}}=0$ and $m=1$. 
Thus, we find that the mother substance of HTS is the AF and that it does not have $\Delta_{\rm{dw}}$. 
This result is consistent with our first assumption that the ground state of the system at zero doping is the AF region. However, $m=1$ indicates that no quantum fluctuation exist, which may have been caused by neglecting the interaction terms that include more than three sites. However, we do not discuss more details about this discrepancy.

Next, we considered when $n^{\delta}\neq0$ and $m\neq0$, which are satisfied by Eqs.\;(\ref{120}) and (\ref{122}). 
Because the integrands of Eqs.\;(\ref{120}) and (\ref{122}) are both positive definite, and $(\cos k_x-\cos k_y)^2 \le 4$, we have:
\begin{align}
1=\text{Eq.\;(\ref{120})} \le \dfrac{4v_{\rm{dw}}}{v_{m}}
\label{125}.
\end{align}
However, we find that the requirement to satisfy Eq.\;(\ref{125}) is $U<4V$, which is not our first assumption of $4V<U$. 
This contradiction has come from the assumption of $m\neq0$ when Eq.\;(\ref{122}) is derived. 
Therefore, we conclude that $m=0$ for any $n^{\delta} \ne0 $ because Eq.\;(\ref{77}) always has trivial solution $m=0$. 
The result $m=0$ for all $n^{\delta} \ne 0$ seems to be consistent with the fact that the experimental magnetic order exists only at a tiny $n^{\delta}$. 

The equations (\ref{120}), (\ref{123}) with $m=0$, and (\ref{117}) are numerically solved in the cases of (a) $U/V=5\;(U/t=6.0, V/t=1.2)$ and (b) $U/V=6\;(U/t=6.3, V/t=1.05)$. 
We choose $v_{\rm{ds}}<0$, which is the condition required for Eq.\;(\ref{117}) to have a solution. 
The values, which are calculated in Eqs.\;(\ref{67-3}) and (\ref{119}), are $v_{\rm{dw}}=2.0333$, $v_{\rm{ds}}=-0.9667$ for (a) and $v_{\rm{dw}}=2.0024$, $v_{\rm{ds}}=-0.8548$ for (b). 

\begin{figure}[b]
\begin{center}
\includegraphics[width=.60\linewidth]{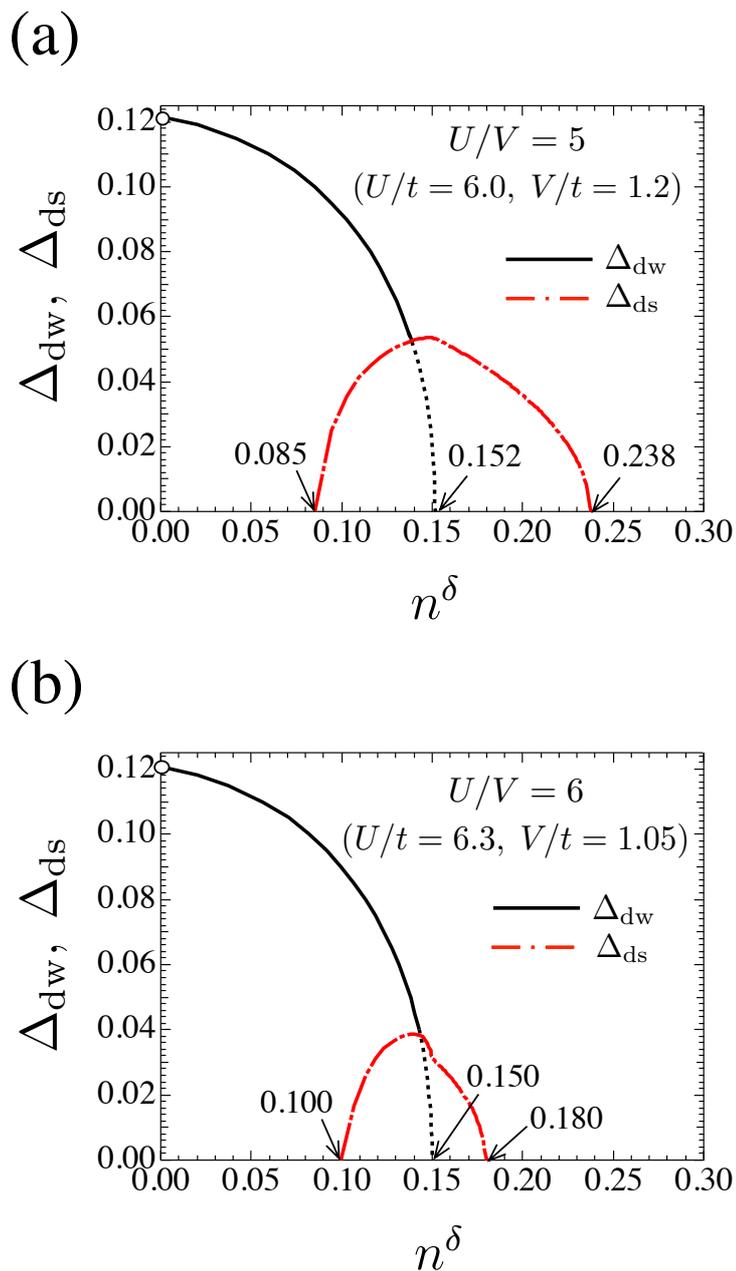}
\end{center}
\caption{Phase diagrams for (a) $U/V=5\;(U/t=6.0, V/t=1.2)$, and for (b) $U/V=6\;(U/t=6.3, V/t=1.05)$. The black and red curves represent $\Delta_{\rm{dw}}$ and $\Delta_{\rm{ds}}$, respectively.}
\label{fig3}
\end{figure}

In FIG.~\ref{fig3}, we show the calculated ground-state phase diagrams of cases (a) and (b), where we set the inverse temperature to $\beta=15$. 
Note that, if the value of $\beta$ is taken as larger, a fine structure appears around the phase boundaries. However, we do not discuss this detail. To focus on investigating qualitatively the structure of the phase diagrams, we regard the obtained results at $\beta=15$ as those of the ground state. 
When $n^{\delta}=0$, we have already obtained $\Delta_{\rm{dw}}=0$ and $m=1$. 
However, as soon as we start doping holes, the DDW order parameters $\Delta_{\rm{dw}}$ emerge and decrease from the maximum value $\Delta_{\rm{dw}}\simeq0.12$ at $n^{\delta}=+0$ for the two cases. 
Thus, $n^{\delta}=0$ is a singular point in our approximation. 
As $n^{\delta}$ increase, $\Delta_{\rm{dw}}$ decrease and vanish at $n^{\delta}\simeq0.152$ and $n^{\delta}\simeq 0.150$ for the two cases, respectively. 
By contrast, the DSC order parameters $\Delta_{\rm{ds}}$ emerge at $n^{\delta}\simeq0.085$ and $0.100$ and vanish at $n^{\delta}\simeq0.238$ and $0.180$ for the two cases, respectively. 
We find that $\Delta_{\rm{ds}}$ has a strong dependence on $|v_{\rm{ds}}|$. 
As $|v_{\rm{ds}}|$ decreases, both the size of the DSC dome and the maximum value of $\Delta_{\rm{ds}}$ decrease.
The maximum values of $\Delta_{\rm{ds}}$ are $\Delta_{\rm{ds}}^{\rm{(max)}}\simeq0.054$ and $0.039$ for the two cases, respectively. 
At $n^{\delta}=0.1358$ and 0.1336, which are near the optimal dopings of the DSC, the van Hove singular point in the density of states coincides with the Fermi level, as described later.

We note that our obtained phase diagrams of FIG.~\ref{fig3} are very similar to the experimental phase diagrams of HTS (see Figure 1b of Ref.\;47) if we assume that the magnitudes of $\Delta_{\rm{dw}}$ and $\Delta_{\rm{ds}}$ are proportional to the transition temperatures. 
We thus conclude that of the pseudogap phase can be interpreted as the DDW phase, which has been proposed by Chakravarty \textit{et al.} 
Our phase diagram shows that a metal-metal quantum transition point exists under the DSC dome. 
Chakravarty \textit{et al.} proposed the same transition.\cite{ref.Chakravarty} 
Furthermore, FIG.~\ref{fig3} shows that the optimal doping is near or slightly to the left of the metal-metal transition point. In the following section, we discuss the details in the case of (a).

To understand further the phase diagram in FIG.~\ref{fig3} (a), we calculate the energies of the normal ($\Delta_{\rm{dw}}=\Delta_{\rm{ds}}=0$), DDW ($\Delta_{\rm{dw}}\neq0$, $\Delta_{\rm{ds}}=0$), and DSC ($\Delta_{\rm{ds}}\neq0$, $\Delta_{\rm{dw}}\neq0$ or $\Delta_{\rm{ds}}\neq0$, $\Delta_{\rm{dw}}=0$) states. 
Substituting $m=0$, $U/t=6.0$, and $V/t=1.2$ into Eqs.\;(\ref{71}), (\ref{72}), and (\ref{96-0}), we obtain: 
\begin{align}
\frac{\langle\mathcal{H}_{\rm{DSC}}^{\rm{MF}}\rangle}{N}\biggl|_{U/V=5}=-\frac{5}{2}\bigr(n^{\delta}\bigl)^2-2\epsilon_{\rm{F}}n^{\delta}-\frac{70}{9}\left\{\Delta_{\rm{dw}}(n^{\delta})\right\}^2-\frac{34}{9}\left\{\Delta_{\rm{dw}}(n^{\delta})\right\}^2
\label{126},
\end{align}
where $\epsilon_{\rm{F}}=\epsilon_{a}/2$ and we have omitted the constant term and set $t=1$ for simplicity. 
The energies of the normal, DDW, and DSC states are written, respectively, as: 
\begin{align}
E_{{\rm{normal}}}^{(U/V=5)}=&-\frac{5}{2}\bigr(n^{\delta}\bigl)^2-2\epsilon_{\rm{F}}(n^{\delta})n^{\delta}
\label{127},\\
E_{\rm{DDW}}^{(U/V=5)}=&-\frac{5}{2}\bigr(n^{\delta}\bigl)^2-2\epsilon_{\rm{F}}(n^{\delta},\Delta_{\rm{dw}})n^{\delta}-\frac{70}{9}\left\{\Delta_{\rm{dw}}(n^{\delta})\right\}^2
\label{128},\\
E_{\rm{DSC}}^{(U/V=5)}=&E_{\rm{DDW}}^{(U/V=5)}-\frac{34}{9}\left\{\Delta_{\rm{dw}}(n^{\delta})\right\}^2
\label{129}.
\end{align}
Note that $\epsilon_{\rm{F}}(n^{\delta})$ here is satisfied by Eq.\;(\ref{123}) with $m=0$ and $\Delta_{\rm{dw}}=0$, and $\epsilon_{\rm{F}}(n^{\delta},\Delta_{\rm{dw}})$ is satisfied by both Eqs.\;(\ref{120}) and (\ref{123}) with $m=0$ and $\Delta_{\rm{dw}}\ge0$.

\begin{figure}[b]
\begin{center}
\includegraphics[width=.65\linewidth]{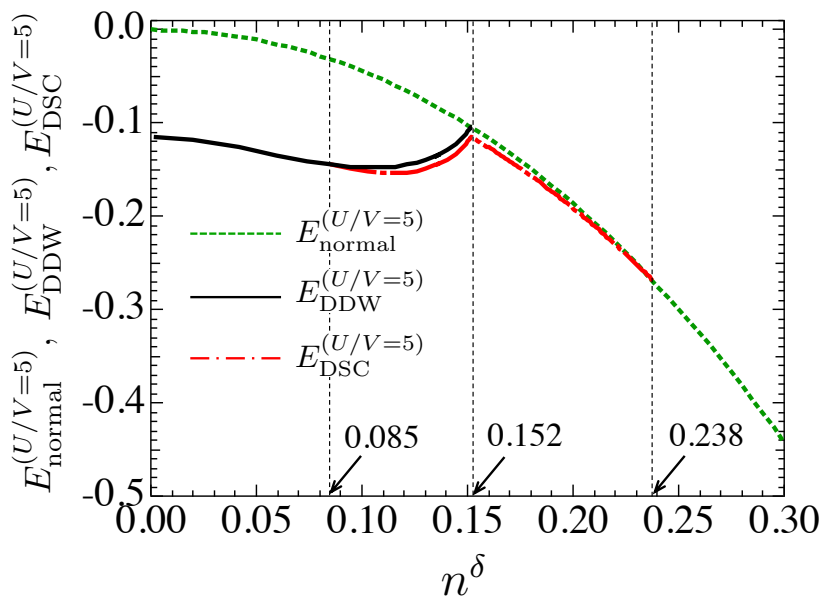}
\end{center}
\caption{Calculated results for the dependence of $E_{\rm{normal}}^{(U/V=5)}$, $E_{\rm{DDW}}^{(U/V=5)}$, and $E_{\rm{DSC}}^{(U/V=5)}$ on $n^{\delta}$. The green, black, and red curves represent $E_{\rm{normal}}^{(U/V=5)}$, $E_{\rm{DDW}}^{(U/V=5)}$, and $E_{\rm{DSC}}^{(U/V=5)}$, respectively.}
\label{fig4}
\end{figure}

The dependences of $E_{\rm{normal}}^{(U/V=5)}$, $E_{\rm{DDW}}^{(U/V=5)}$, and $E_{\rm{DSC}}^{(U/V=5)}$ on $n^{\delta}$ are shown in FIG.~\ref{fig4}. 
In the region of $0<n^{\delta}\lesssim0.085$, the ground state is the DDW state. 
When $n^{\delta}\simeq0.085$, the DSC emerges and, in the region of $0.085\lesssim n^{\delta}\lesssim0.152$, the DSC state coexists with the DDW. 
At $n^{\delta}\simeq0.152$, the DDW vanishes and, in the region of $0.152\lesssim n^{\delta}\lesssim0.238$, 
the ground state becomes the DSC state without the DDW.
Furthermore, at $n^{\delta}\simeq0.238$, the DSC vanishes and, in the region of $0.238\lesssim n^{\delta}$, the normal state becomes the ground state. 

As shown in FIG.~\ref{fig4}, in the underdoped region, the energy difference between $E_{\rm{DDW}}^{(U/V=5)}$ and $E_{\rm{normal}}^{(U/V=5)}$ is so large that the DDW order can be detected by the Monte Carlo simulation. 

\subsection{Fermi surface in the $d_{x^2-y^2}$-wave density wave state}

\begin{figure}[b]
\begin{center}
\includegraphics[width=.96\linewidth]{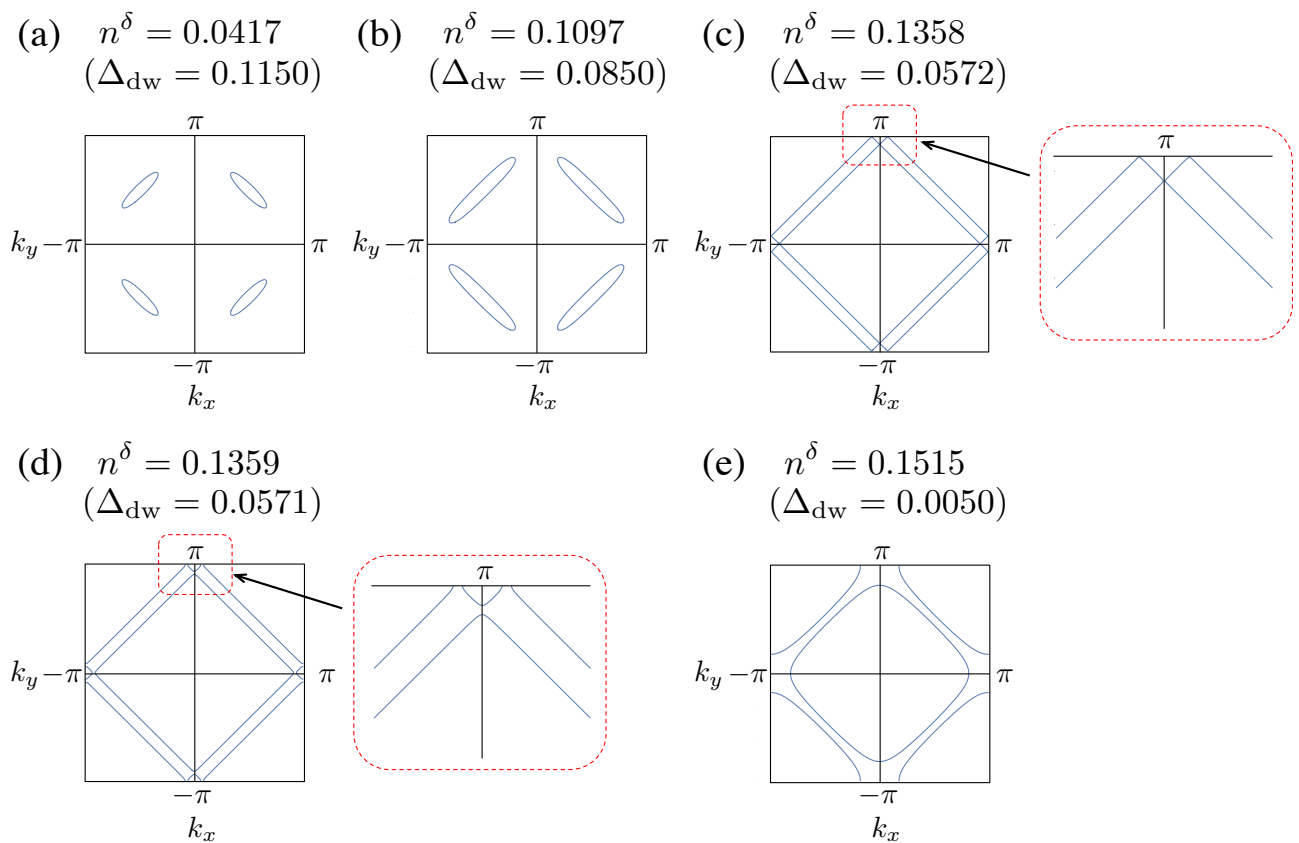}
\end{center}
\caption{Fermi surfaces in the DDW state when the hole concentration $n^{\delta}$ (DDW order parameters $\Delta_{\rm{dw}}$) are: (a) $n^{\delta}=0.0417$ ($\Delta_{\rm{dw}}=0.1150$), (b) $n^{\delta}=0.1097$ ($\Delta_{\rm{dw}}=0.0850$), (c) $n^{\delta}=0.1358$ ($\Delta_{\rm{dw}}=0.0572$), (d) $n^{\delta}=0.1359$ ($\Delta_{\rm{dw}}=0.0571$), and (e) $n^{\delta}=0.1515$ ($\Delta_{\rm{dw}}=0.0050$), when $U/V=5\;(U/t=6.0, V/t=1.2)$. The insets in FIGs. (c) and (d) show the enlarged plots around the points $(k_x,k_y)=(0, \pi)$.
}
\label{fig5}
\end{figure}

In FIG.~\ref{fig5}, we show the Fermi surface in the DDW state when the hole concentrations $n^{\delta}$ (DDW order parameters $\Delta_{\rm{dw}}$) are: (a) $n^{\delta}=0.0417$ ($\Delta_{\rm{dw}}=0.1150$), (b) $n^{\delta}=0.1097$ ($\Delta_{\rm{dw}}=0.0850$), (c) $n^{\delta}=0.1358$ ($\Delta_{\rm{dw}}=0.0572$), (d) $n^{\delta}=0.1359$ ($\Delta_{\rm{dw}}=0.0571$), and (e) $n^{\delta}=0.1515$ ($\Delta_{\rm{dw}}=0.0050$), when $U/V=5\;(U/t=6.0, V/t=1.2)$. 

When $\Delta_{\rm{dw}}$ is large ($n^{\delta}$ is small), the Fermi surfaces are small ellipses, and as $\Delta_{\rm{dw}}$ becomes small ($n^{\delta}$ is large), they expand to large Fermi surfaces (see FIGs.~\ref{fig5}(a)$\to$(b)$\to$(c)). 
As shown in FIG.~\ref{fig5}(c), at $n^{\delta}=0.1358$ ($\Delta_{\rm{dw}}=0.0572$), four Fermi surfaces are in contact with each other near $(k_x,k_y)=(0, \pm\pi)$ and $(\pm\pi, 0)$. 
The inset in FIG.~\ref{fig5}(c) shows that the Fermi surfaces of the first and second quadrants are in contact with each other near $(k_x,k_y)=(0, \pi)$. At $n^{\delta}=0.1358$, where the topology of the Fermi surface changes, the van Hove singular point in the density of states coincides with the Fermi level. 
In addition, as shown in FIG.~\ref{fig5}(d), immediately after the four Fermi surfaces come into contact, they are connected to each other and produce small pockets at $(k_x,k_y)=(0, \pm\pi)$ and $(\pm\pi, 0)$. 
FIG.~\ref{fig5}(e) shows that the united large Fermi surface emerges for more than $n^{\delta}\simeq 0.1359$. 

\subsection{Density of states in the $d_{x^2-y^2}$-wave density wave state}

The density of states per site in the DDW state is defined as: 
\begin{align}
g(\omega)=&\frac{1}{4\pi^2}\int_{-\pi}^{\pi}\int_{-\pi}^{\pi}dk_xdk_y\delta(\omega-\epsilon_{\bm{k}})
\label{130}, \\
\epsilon_{\bm{k}}=&2\sqrt{(\cos k_x+\cos k_y)^2+p^2(\cos k_x-\cos k_y)^2}-\epsilon_a
\label{131}
\end{align} 
where $\epsilon_{\bm{k}}(=-E^-_{\bm{k},1,m=0})$ is the dispersion relation of the quasi-particle (see Eq.\;(\ref{78-2})), $p=v_{\rm{dw}}\Delta_{\rm{dw}}$, and we set $t=1$. 
From Eq.\;(\ref{e4}) of Appendix E, the density of states must be calculated on the two regions depending on $\omega$. 

When $-\epsilon_a\le\omega<\frac{4p}{\sqrt{1+p^2}}-\epsilon_a$, we obtain: 
\begin{align}
g(\omega)=&\frac{\omega+\epsilon_a}{4\pi^2p}\int_{\phi}^{\pi-\phi}dk_y\frac{1}{\sin k_x^0\sqrt{\cos^2\phi-\cos^2k_y}}
\label{132},
\end{align} 
where 
\begin{align}
k_x^0=&\cos^{-1}\left\{-\frac{1-p^2}{1+p^2}\cos k_y+\frac{2p}{1+p^2}\sqrt{\cos^2\phi-\cos^2k_y}\right\}
\label{133},\\
\phi=&\cos^{-1}\left\{\frac{\sqrt{1+p^2}}{4p}(\omega+\epsilon_a)\right\}
\label{134}.
\end{align} 
On the other hand, when $\frac{4p}{\sqrt{1+p^2}}-\epsilon_a\le\omega<4-\epsilon_a$, we obtain: 
\begin{align}
g(\omega)=&\frac{\omega+\epsilon_a}{4\pi^2p}\int_{0}^{\phi}dk_y\frac{1}{\sin k_x^0\sqrt{\frac{1+p^2}{16p^2}(\omega+\epsilon_a)^2-\cos^2k_y}}
\label{135},
\end{align} 
where 
\begin{align}
k_x^0=&\cos^{-1}\left\{-\frac{1-p^2}{1+p^2}\cos k_y+\frac{2p}{1+p^2}\sqrt{\frac{1+p^2}{16p^2}(\omega+\epsilon_a)^2-\cos^2k_y}\right\}
\label{136},\\
\phi=&\cos^{-1}\left\{-\frac{1-p^2}{1+p^2}+\frac{2p}{1+p^2}\sqrt{\frac{1+p^2}{16p^2}(\omega+\epsilon_a)^2-1}\right\}
\label{137}.
\end{align} 

\begin{figure}[b]
\begin{center}
\includegraphics[width=.41\linewidth]{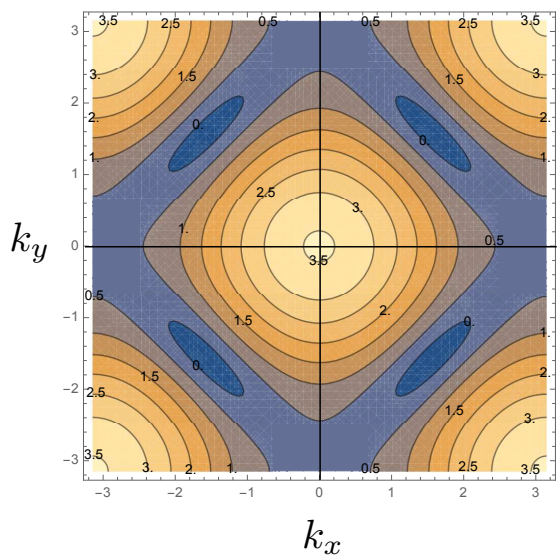}
\end{center}
\caption{Equal-energy surface of Eq.\;(\ref{131}) at $n^{\delta}=0.0417$ ($\Delta_{\rm{dw}}=0.1150$, and $\epsilon_a=0.4538$). }
\label{fig6}
\end{figure}

In FIG.~\ref{fig6}, we show the equal energy surface of Eq.\;(\ref{131}) at $n^{\delta}=0.0417$ ($\Delta_{\rm{dw}}=0.1150$ and $\epsilon_a=0.4538$).
In the case of $\omega\le\frac{4p}{\sqrt{1+p^2}}-\epsilon_a$, for example, when we limit ourselves only to the first quadrant, the contour solution, which is denoted by a small ellipse in FIG.~\ref{fig6}, occupies the region of $\phi(\omega)\le k_x\le \pi-\phi(\omega)$ and $\phi(\omega)\le k_y\le \pi-\phi(\omega)$ from Eqs.\;(\ref{132}) and (\ref{134}). 
Similarly, in the case of $\omega\ge\frac{4p}{\sqrt{1+p^2}}-\epsilon_a$, we have a large contour solution in the region of $0\le k_x\le\phi(\omega)$ and $0\le k_y\le\phi(\omega)$ from Eqs.\;(\ref{135}) and (\ref{137}). It should be noted that the topology of the contour line changes at $\omega=\frac{4p}{\sqrt{1+p^2}}-\epsilon_a$.

\begin{figure}[t]
\begin{center}
\includegraphics[width=.65\linewidth]{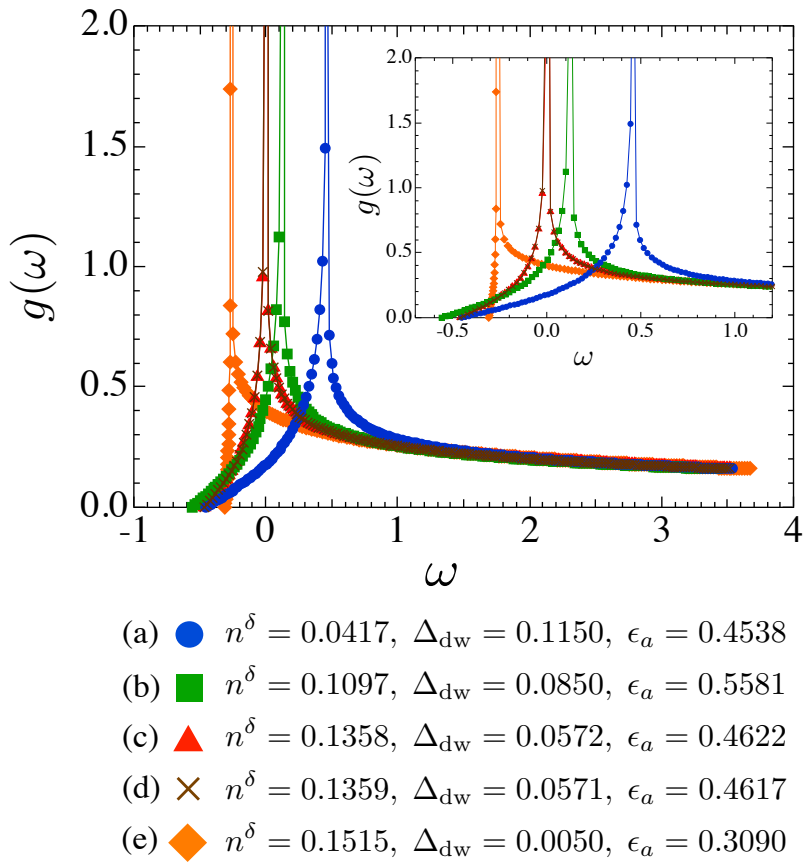}
\end{center}
\caption{Calculation results for the dependence of the density of states in the DDW state $g(\omega)$ on $\omega$ when: (a) $n^{\delta}=0.0417$, $\Delta_{\rm{dw}}=0.1150$, and $\epsilon_a=0.4538$, (b) $n^{\delta}=0.1097$, $\Delta_{\rm{dw}}=0.0850$, and $\epsilon_a=0.5581$, (c) $n^{\delta}=0.1358$, $\Delta_{\rm{dw}}=0.0572$, and $\epsilon_a=0.4622$, (d) $n^{\delta}=0.1359$, $\Delta_{\rm{dw}}=0.0571$, and $\epsilon_a=0.4617$, and (e) $n^{\delta}=0.1515$, $\Delta_{\rm{dw}}=0.0050$, and $\epsilon_a=0.3401$, when $U/V=5\;(U/t=6.0, V/t=1.2)$. The inset shows the area surrounding the region of $\omega=\omega_0$, which is the point of the van Hove singularity. }
\label{fig7}
\end{figure}

In FIG.~\ref{fig7}, we show the dependence of $g(\omega)$ on $\omega$ when: (a) $n^{\delta}=0.0417$, $\Delta_{\rm{dw}}=0.1150$, and $\epsilon_a=0.4538$, (b) $n^{\delta}=0.1097$, $\Delta_{\rm{dw}}=0.0850$, and $\epsilon_a=0.5581$, (c) $n^{\delta}=0.1358$, $\Delta_{\rm{dw}}=0.0572$, and $\epsilon_a=0.4622$, (d) $n^{\delta}=0.1359$, $\Delta_{\rm{dw}}=0.0571$, and $\epsilon_a=0.4617$, and (e) $n^{\delta}=0.1515$, $\Delta_{\rm{dw}}=0.0050$, and $\epsilon_a=0.3090$, when $U/V=5\;(U/t=6.0, V/t=1.2)$. 
These figures show that in all cases, $g(\omega)$ have van Hove singularities at $\omega_0=\frac{4p}{\sqrt{1+p^2}}-\epsilon_a$ with $p=v_{\rm{dw}}\Delta_{\rm{dw}}$, where $\omega_0=0.4570$, 0.1231, 0, $-0.0004$, and $-0.2683$, respectively, and $g(\omega)=0$ at $\omega\le-\epsilon_a$. 
These facts indicate that we can obtain the value of $p=v_{\rm{dw}}\Delta_{\rm{dw}}$ if we experimentally observe the density of states to extract the lower boundary $\omega=-\epsilon_a$ and the position of the van Hove singularity $\omega_0$.

\section{Summary}

In this study, the extended Hubbard Hamiltonian with NN Coulomb repulsion on the square lattice was studied by applying the hole picture to the AF region to obtain the phase diagrams of the DDW and DSC phases. 
The phase diagrams, which were obtained by regarding the order parameter as the transition temperature, were qualitatively consistent with the phase diagrams observed in experiments of the HTS.\cite{ref.experiment} 
By overlooking both phase diagrams, we conclude that the pseudogap can be interpreted as the DDW order parameter. 
 
In the underdoped regions, the DDW order parameter decreased with increasing hole doping and vanished at the new metal-metal quantum critical transition point under the DSC dome, which was predicted by Chakravarty \textit{et al.}\cite{ref.Chakravarty} The DSC states were established based on the DDW state. 
As the hole doping increased, the shape of the Fermi surface with the DDW order changed from four small to large ellipses. When the topology of the Fermi surface changed, the van Hove singular point in the density of states was located at the Fermi energy. 
In addition, we stress that we could obtain the theoretical value of $p=v_{\rm{dw}}\Delta_{\rm{dw}}$ if we experimentally observed both the value of energy at which the density of states appeared and the value of the van Hove singular point, as these are written by $\omega=-\epsilon_a$ and $\omega_0=\frac{4p}{\sqrt{1+p^2}}-\epsilon_a$, respectively.
By contrast, in the overdoped region, the DDW collapsed and the system became a normal metal and the Fermi surface became the large contours. 
The DSC was suppressed simultaneously. 
The ground state was as follows: the AF region was found only at $n^{\delta}=0$ in our approximation and as $n^{\delta}$ increased, the ground state became the DDW, DDW+DSC, DSC, and normal metal states. 

We conclude that the hidden order proposed by Chakravarty \textit{et al.}\cite{ref.Chakravarty} can be obtained from our effective Hamiltonian by using the canonical transformation. 
We expect that our obtained results assist in the qualitative explanation of the experimental results if the DDW order is present. 

\vskip 0.5cm\noindent
	{\bf Acknowledgment}
	
This work was supported by JSPS KAKENHI Grant Numbers JP17J05190 and JP17K05519.

\appendix
\renewcommand{\theequation}{A.\arabic{equation}}
\setcounter{equation}{0}
\section{Calculation details for the operator $S$}

We can obtain $S$ starting from Eq.\;(\ref{9}) as follows. 
By using the eigenstate $| m \rangle $ of $H_0$, which satisfies $H_0|m\rangle=E_m|m\rangle$, we obtain:
\begin{equation}
\langle l|S|m\rangle=\dfrac{\langle l|H_1(1)|m\rangle}{E_l-E_m}=-i\int_0^{\infty}\langle l|e^{i H_0\lambda}H_1(1)e^{-i H_0\lambda}|m\rangle d \lambda
\label{a1}.
\end{equation}
Therefore, from Eq.\;(\ref{a1}), the operator equation is obtained by:
\begin{equation}
S=-i\int_0^{\infty}d\lambda e^{iH_0\lambda }H_1(1)e^{-iH_0\lambda}
\label{a2}, 
\end{equation}
which is same as Eq.\;(\ref{10}). 
Next, substituting Eq.\;(\ref{7}) into Eq.\;(\ref{a2}) and using the commutation relations and $e^{iH_0\lambda}c_{i\sigma}e^{-iH_0\lambda}=e^{-i\lambda X_{i\sigma}}c_{i,\sigma}$, where $X_{i,\sigma}=Un_{i,-\sigma}+V\displaystyle{\sum_{\eta}n_{i+\eta}}$ and $\eta$ is the NN vector, we obtain:
\begin{align}
S=&it\sum_{i\in{\rm{A,B}}}\sum_{\delta,\sigma}\int_0^{\infty}d\lambda\Bigl[\left\{e^{i\lambda(U-V)}n_{i-\sigma}(1-n_{i+\delta,-\sigma}) 
+e^{-i\lambda(U-V)}(1-n_{i,-\sigma})n_{i+\delta,-\sigma}\right\} \notag \\
&\qquad\qquad\quad\times e^{ i\lambda V\sum_{\eta(\neq\delta)}(n_{i+\eta}-n_{i+\delta -\eta})}c_{i,\sigma}^{\dagger}c_{i+\delta,\sigma }+\text{h.c} \Bigr]\notag \\
=&it\sum_{i\in{\rm{A}}}\sum_{\delta}\int_0^{\infty}d\lambda \Bigl[\left\{n_i^a n_{i+\delta}^de^{i\lambda(U-V)}+(1-n_i^a)(1-n_{i+\delta}^d)e^{-i\lambda(U-V)}\right\} \notag \\
&\qquad\qquad\quad\quad\times e^{i \lambda V\sum_{\eta(\neq\delta)}(n_{i+\eta}^c-n_{i+\eta}^d+n_{i+\delta-\eta}^b-n_{i+\delta-\eta}^a)}b_ic_{i+\delta}\notag \\
&\qquad\qquad\quad+\left\{(1-n_i^b)(1-n_{i+\delta}^c)e^{i\lambda(U-V)}+n_i^bn_{i+\delta}^ce^{-i\lambda(U-V)}\right\} \notag \\
&\qquad\qquad\quad\quad\times e^{i \lambda V\sum_{\eta(\neq\delta)}(n_{i+\eta}^c-n_{i+\eta}^d+n_{i+\delta-\eta}^b-n_{i+\delta-\eta}^a) }a_i^{\dagger}d_{i+\delta}^{\dagger}\Bigr]- \text{h.c.}
\label{a3}.
\end{align}
Furthermore, substituting the following relation,
\begin{equation}
\exp\Bigl(\pm i\lambda V\sum_{i}n_i\Bigr)=\prod_i\left\{1+(e^{\pm i\lambda V}-1)n_i\right\}
\label{a4}, 
\end{equation}
into Eq.\;(\ref{a3}), while neglecting terms consisting of a product of more than four operators and integrating $\lambda$, we finally obtain:
\begin{align}
S=&t\sum_{i,\delta}\biggl\{\frac{1}{U-V}(1-n_i^a-n_{i+\delta}^d)+\sum_{\eta(\neq \delta)}\left(\frac{1}{U-2V}-\frac{1}{U-V}\right)(n_{i+\eta}^c+n_{i+\delta-\eta}^b)  \notag \\
&\qquad+\sum_{\eta(\neq \delta)}\left(\frac{1}{U}-\dfrac{1}{U-V}\right)(n_{i+\eta}^d+n_{i+\delta-\eta}^a)\biggr\} (b_i c_{i+\delta}-c_{i+\delta}^{\dagger}b_i^{\dagger})\notag \\
-&t\sum_{i,\delta}\biggl\{\frac{1}{U-V}(1-n_i^b-n_{i+\delta}^c)+\sum_{\eta(\neq \delta)}\left(\frac{1}{U-2V}-\dfrac{1}{U-V}\right)(n_{i+\eta}^d+n_{i+\delta-\eta}^a) \notag \\
&\qquad+\sum_{\eta(\neq \delta)}\left(\frac{1}{U}-\dfrac{1}{U-V}\right)(n_{i+\eta}^c+n_{i+\delta-\eta}^b)\biggr\} (a_i d_{i+\delta}-d_{i+\delta}^{\dagger}a_i^{\dagger})
\label{a5}.
\end{align}

\renewcommand{\theequation}{B.\arabic{equation}}
\setcounter{equation}{0}
\section{Calculation details for the effective Hamiltonian $\mathcal{H}_{eff}$}

We calculate  Eq.\;(\ref{20}) and obtain Eq.\;(\ref{21}) using the approximations (I) and (II). 
The quantities neglected in our calculation are denoted by $o(n)$.

First, we calculate $[S,H_0]$ using:
\begin{align}
[S,H_0]=&\sum_{i,\delta}I_0\bigl(1-N_{i,i+\delta}^{ad}\bigr)[F_{i,i+\delta}^{bc}, H_0]-\sum_{i,\delta}I_0\bigl(1-N_{i,i+\delta}^{bc}\bigr)[F_{i,i+\delta}^{ad}, H_0]+o(n)
\label{b1},
\end{align}
where 
\begin{align}
[F_{i,i+\delta}^{bc}, H_0]=&(U-V)\bigl(1-N_{i,i+\delta}^{ad}\bigr)\bigl(b_i c_{i+\delta}+c_{i+\delta}^{\dagger}b_{i}^{\dagger}\bigr)+o(n)
\label{b2},\\
[F_{i,i+\delta}^{ad}, H_0]=&(U-V)\bigl(1-N_{i,i+\delta}^{bc}\bigr)\bigl(a_i d_{i+\delta}+d_{i+\delta}^{\dagger}a_{i}^{\dagger}\bigr)+o(n)
\label{b3}.
\end{align}
Then, substituting Eqs.\;(\ref{b2}) and (\ref{b3}) into Eq.\;(\ref{b1}), we obtain:
\begin{align}
[S, H_0]=t\sum_{i,\delta}\left\{\bigl(1-N_{i,i+\delta}^{ad}\bigr)\bigl(b_i c_{i+\delta}+c_{i+\delta}^{\dagger}b_i^{\dagger}\bigr)-\bigl(1-N_{i,i+\delta}^{bc}\bigr)\bigl(a_i d_{i+\delta}+d_{i+\delta}^{\dagger}a_i^{\dagger}\bigr)\right\}+o(n)
\label{b4}.
\end{align}
Furthermore, from Eqs.\;(\ref{4}) and (\ref{b4}), we have:
\begin{align}
H_1+[S, H_0]=-t\sum_{i,\delta}\left\{N_{i,i+\delta}^{ad}\bigl(b_i c_{i+\delta}+c_{i+\delta}^{\dagger}b_i^{\dagger}\bigr)-N_{i,i+\delta}^{bc}\bigl(a_i d_{i+\delta}+d_{i+\delta}^{\dagger}a_i^{\dagger}\bigr)\right\}+o(n)
\label{b5}.
\end{align}

Next, we calculate $[S,H_1]$ using:
\begin{align}
[S,H_1]=&I_0\tilde{I}_0+I_1\tilde{I}_1+I_2\tilde{I}_2
\label{b6},
\end{align}
where
\begin{align}
\tilde{I}_0=&\sum_{i,\delta}\left[\bigl(1-N_{i,i+\delta}^{ad}\bigr)F_{i,i+\delta}^{bc}-\bigl(1-N_{i,i+\delta}^{bc}\bigr)F_{i,i+\delta}^{ad},H_1\right]\notag \\
	   =&-8Nt+16t\sum_i\bigl(n_i^a+n_i^b\bigr)+16t\sum_l\bigl(n_l^c +n_l^d\bigr)-16t\sum_in_i^a n_i^b-16t\sum_l n_l^c n_l^d \notag \\
&-4t\sum_{i,\delta}\bigl(n_i^a n_{i+\delta}^c+n_i^b n_{i+\delta}^d\bigr)+2t\sum_{i,\delta}F_{i,i+\delta}^{ad}F_{i,i+\delta}^{bc}+2t\sum_{l,\delta}F_{l+\delta,l}^{ad}F_{l+\delta,l}^{bc}+o(n)
\label{b7},\\
\tilde{I}_1=&\sum_{i,\delta}\sum_{\eta(\neq\delta)}\left[N_{i+\delta-\eta,i+\eta}^{bc}F_{i,i+\delta}^{bc}-N_{i+\delta-\eta,i+\eta}^{ad}F_{i,i+\delta}^{ad},H_1\right]\notag \\
	   =&-24t\sum_i\left(n_i^a+n_i^b\right)-24t\sum_l\left(n_l^c+n_l^d\right)+12t\sum_{i,\delta}\left(n_i^b n_{i+\delta}^c+n_i^a n_{i+\delta}^d\right)+o(n)
\label{b8},\\
\tilde{I}_2=&\sum_{i,\delta}\sum_{\eta(\neq\delta)}\left[N_{i+\delta-\eta,i+\eta}^{ad}F_{i,i+\delta}^{bc}-N_{i+\delta-\eta,i+\eta}^{bc}F_{i,i+\delta}^{ad},H_1\right]\notag \\
	   =&-24t\sum_i\left(n_i^a+n_i^b\right)-24t\sum_l\left(n_l^c+n_l^d\right)+12t\sum_{i,\delta}\left(n_i^b n_{i+\delta}^d+n_i^a n_{i+\delta}^c\right)+o(n)
\label{b9}.
\end{align}

Then, substituting Eqs.\;(\ref{3}) and (\ref{b6}) into Eq.\;(\ref{20}), we finally obtain Eq.\;(\ref{21}).

\renewcommand{\theequation}{C.\arabic{equation}}
\setcounter{equation}{0}
\section{Derivation of the self-consistent equations for $\Delta$ and $\Delta^{\prime}$}

Starting from Eq.\;(\ref{52}), we derive the self-consistent Eqs.\;(\ref{60}) and (\ref{61}) and solve these as follows. 

Using Eq.\;(\ref{52}), the Heisenberg equations for $a_{\bm{k}}$, $d_{\bm{k}}^{\dagger}$, $b_{\bm{k}}$, and $c_{\bm{k}}^{\dagger}$ are written as: 
\begin{align}
i\hbar\frac{\partial a_{\bm{k}}}{\partial t}=&[a_{\bm{k}}, \mathcal{H}_{eff}^{\rm{MF}}]=\epsilon_a a_{\bm{k}}-\Gamma_{\bm{k}} d_{\bm{k}}^{\dagger} 
\label{c1},\\
i\hbar\frac{\partial d_{\bm{k}}^{\dagger}}{\partial t}=&[d_{\bm{k}}^{\dagger}, \mathcal{H}_{eff}^{\rm{MF}}]=-\epsilon_b d_{\bm{k}}^{\dagger}-\Gamma_{\bm{k}}^{\dagger} a_{\bm{k}} 
\label{c2},\\
i\hbar\frac{\partial b_{\bm{k}}}{\partial t}=&[b_{\bm{k}}, \mathcal{H}_{eff}^{\rm{MF}}]=\epsilon_b b_{\bm{k}}+\Gamma_{\bm{k}}^{\prime}c_{\bm{k}}^{\dagger} 
\label{c3},\\
i\hbar\frac{\partial c_{\bm{k}}^{\dagger}}{\partial t}=&[c_{\bm{k}}^{\dagger}, \mathcal{H}_{eff}^{\rm{MF}}]=-\epsilon_a c_{\bm{k}}^{\dagger}+\Gamma_{\bm{k}}^{\prime\dagger}b_{\bm{k}} 
\label{c4}. 
\end{align}
Then, from Eqs.\;(\ref{c1}) and (\ref{c2}), we obtain:
\begin{align}
(\omega-\epsilon_a)\ll a_{\bm{k}},a_{\bm{k}}^{\dagger}\gg_{\omega}+\Gamma_{\bm{k}}\ll d_{\bm{k}}^{\dagger},a_{\bm{k}}^{\dagger}\gg_{\omega}=&1
\label{c5}, \\
\Gamma_{\bm{k}}^{\dagger}\ll a_{\bm{k}},a_{\bm{k}}^{\dagger}\gg_{\omega} +(\omega+\epsilon_b)\ll d_{\bm{k}}^{\dagger},a_{\bm{k}}^{\dagger}\gg_{\omega}=&0
\label{c6},
\end{align}
where $\ll A,B\gg_{\omega}$ is the Fourier component of the double time Green function, which is written as:
\begin{align}
\ll A,B\gg_{\omega}=\int_{-\infty}^{\infty}\ll A(t),B\gg e^{i\omega t}dt
\label{c7},\\
\ll A(t),B\gg=-i\theta(t)\langle\left[A(t),B\right]\rangle
\label{c8}.
\end{align}
From Eqs.\;(\ref{c5}) and (\ref{c6}), we obtain:
\begin{align}
\ll a_{\bm{k}},a_{\bm{k}}^{\dagger}\gg_{\omega}=\dfrac{1}{E_{\bm{k}}^{+}-E_{\bm{k}}^{-}}\biggl(\dfrac{E_{\bm{k}}^{+}+\epsilon_b}{\omega-E_{\bm{k}}^{+}}-\dfrac{E_{\bm{k}}^{-}+\epsilon_b}{\omega-E_{\bm{k}}^{-}}\biggr)
\label{c9}, \\
\ll d_{\bm{k}}^{\dagger},a_{\bm{k}}^{\dagger}\gg_{\omega}=\dfrac{-\Gamma_{\bm{k}}^{\dagger}}{E_{\bm{k}}^{+}-E_{\bm{k}}^{-}}\biggl(\dfrac{1}{\omega-E_{\bm{k}}^{+}}-\dfrac{1}{\omega-E_{\bm{k}}^{-}}\biggr)
\label{c10},
\end{align}
where $E_{\bm{k}}^{\pm}$ is defined by Eq.\;(\ref{62}). 
Furthermore, using the spectral theorem, we obtain:
\begin{align}
\langle a_{\bm{k}}^{\dagger}a_{\bm{k}} \rangle =&\dfrac{1}{E_{\bm{k}}^{+}-E_{\bm{k}}^{-}}\Big\{(E_{\bm{k}}^{+}+\epsilon_b)f(E_{\bm{k}}^{+})-(E_{\bm{k}}^{-}+\epsilon_b)f(E_{\bm{k}}^{-})\Big\}
\label{c12}, \\
\langle a_{\bm{k}}^{\dagger}d_{\bm{k}}^{\dagger}\rangle =&\dfrac{-\Gamma_{\bm{k}}^{\dagger}}{E_{\bm{k}}^{+}-E_{\bm{k}}^{-}}\left\{f(E_{\bm{k}}^{+})-f(E_{\bm{k}}^{-})\right\}
\label{c13},
\end{align}
where $f(E)=\frac{1}{e^{\beta E}+1}$ is the Fermi distribution function. 
Then, by using Eq.\;(\ref{c13}), self-consistent Eq.\;(\ref{60}) is obtained by:
\begin{align}
\Delta=&\dfrac{1}{2N}\sum_i \bigl(\langle a_id_{i+\bm{e}_x}\rangle+\langle a_id_{i-\bm{e}_x}\rangle-\langle a_id_{i+\bm{e}_y}\rangle-\langle a_id_{i-\bm{e}_y} \rangle\bigr) \nonumber \\
=&\dfrac{1}{N}\sum_{\bm{k}} (\cos k_x-\cos k_y)\langle a_{\bm{k}}d_{\bm{k}} \rangle \nonumber \\
=&\dfrac{1}{2 N}\sum_{\bm{k}} \dfrac{s_{\bm{k}}^2\left\{V^{\prime\prime}\Delta+4tI_0(\Delta^{\prime}-\Delta^{\prime\dagger})\right\}}{\sqrt{(\epsilon_a+\epsilon_b)^2+4|\Gamma_{\bm{k}}|^2}}\left\{f(E_{\bm{k}}^{-})-f(E_{\bm{k}}^{+})\right\}
\label{c14},
\end{align}
where we used Eqs.\;(\ref{53}) and (\ref{56}).

On the other hand, from Eqs.\;(\ref{c3}) and (\ref{c4}), we obtain:
\begin{align}
(\omega-\epsilon_b)\ll b_{\bm{k}},b_{\bm{k}}^{\dagger}\gg_{\omega}-\Gamma_{\bm{k}}^{\prime}\ll c_{\bm{k}}^{\dagger},b_{\bm{k}}^{\dagger} \gg_{\omega}=&1
\label{c15},\\
-\Gamma_{\bm{k}}^{\prime\dagger}\ll b_{\bm{k}},b_{\bm{k}}^{\dagger}\gg_{\omega} +(\omega+\epsilon_a)\ll c_{\bm{k}}^{\dagger},b_{\bm{k}}^{\dagger}\gg_{\omega}=&0
\label{c16}. 
\end{align}
Then, from Eqs.\;(\ref{c15}) and (\ref{c16}), we obtain:
\begin{align}
\ll b_{\bm{k}},b_{\bm{k}}^{\dagger}\gg_{\omega}=&\dfrac{1}{E_{\bm{k}}^{\prime+}-E_{\bm{k}}^{\prime-}}\biggl(\dfrac{E_{\bm{k}}^{\prime+}+\epsilon_a}{\omega-E_{\bm{k}}^{\prime+}}-\dfrac{E_{\bm{k}}^{\prime-}+\epsilon_a}{\omega-E_{\bm{k}}^{\prime-}}\biggr)
\label{c17},\\
\ll c_{\bm{k}}^{\dagger}, b_{\bm{k}}^{\dagger}\gg_{\omega}=&\dfrac{\Gamma_{\bm{k}}^{\prime\dagger}}{E_{\bm{k}}^{\prime+}-E_{\bm{k}}^{\prime-}}\biggl(\dfrac{1}{\omega-E_{\bm{k}}^{\prime+}}-\dfrac{1}{\omega-E_{\bm{k}}^{\prime-}}\biggr)
\label{c18},
\end{align}
where $E_{\bm{k}}^{\prime\pm}$ is defined by Eq.\;(\ref{63}). 
We then obtain:
\begin{align}
\langle b_{\bm{k}}^{\dagger}b_{\bm{k}} \rangle=&\dfrac{1}{E_{\bm{k}}^{\prime+}-E_{\bm{k}}^{\prime-}}\Big\{(E_{\bm{k}}^{\prime+}+\epsilon_a)f(E_{\bm{k}}^{\prime+})-(E_{\bm{k}}^{\prime-}+\epsilon_a)f(E_{\bm{k}}^{\prime-})\Big\}
\label{c20}, \\
\langle b_{\bm{k}}^{\dagger}c_{\bm{k}}^{\dagger}\rangle=&\dfrac{\Gamma_{\bm{k}}^{\prime\dagger}}{E_{\bm{k}}^{\prime+}-E_{\bm{k}}^{\prime-}}\left\{f(E_{\bm{k}}^{\prime+})-f(E_{\bm{k}}^{\prime-})\right\}
\label{c21}.
\end{align}
Then, by using Eq.\;(\ref{c21}), we obtain self-consistent Eq.\;(\ref{61}) by:
\begin{align}
\Delta^{\prime}=&\dfrac{1}{2N}\sum_i \bigl(\langle b_ic_{i+\bm{e}_x}\rangle+\langle b_ic_{i-\bm{e}_x}\rangle-\langle b_ic_{i+\bm{e}_y}\rangle-\langle b_ic_{i-\bm{e}_y} \rangle\bigr) \nonumber \\
=&\dfrac{1}{N}\sum_{\bm{k}} (\cos k_x-\cos k_y)\langle b_{\bm{k}}c_{\bm{k}} \rangle \nonumber \\
=&\dfrac{1}{2 N}\sum_{\bm{k}} \dfrac{s_{\bm{k}}^2\left\{V^{\prime\prime}\Delta^{\prime}+4tI_0(\Delta-\Delta^{\dagger})\right\}}{\sqrt{(\epsilon_a+\epsilon_b)^2+4|\Gamma^{\prime}_{\bm{k}}|^2}}\left\{f(E_{\bm{k}}^{\prime-})-f(E_{\bm{k}}^{\prime+})\right\}
\label{c22}.
\end{align}

We define $R, K, R^{\prime} $ and $K^{\prime}$ as: 
\begin{align}
R=&\dfrac{1}{2N}\sum_{\bm{k}}\dfrac{s_{\bm{k}}^2V^{\prime\prime}\left\{f(E_{\bm{k}}^-)-f(E_{\bm{k}}^+)\right\}}{\sqrt{(\epsilon_a+\epsilon_b)^2+4|\Gamma_{\bm{k}}|^2}}
\label{c23},\\
K=&\dfrac{1}{2N}\sum_{\bm{k}}\dfrac{4s_{\bm{k}}^2tI_0\left\{f(E_{\bm{k}}^-)-f(E_{\bm{k}}^+)\right\}}{\sqrt{(\epsilon_a+\epsilon_b)^2+4|\Gamma_{\bm{k}}|^2}} 
\label{c24},\\
R^{\prime}=&\dfrac{1}{2N}\sum_{\bm{k}}\dfrac{s_{\bm{k}}^2V^{\prime\prime}\left\{f(E_{\bm{k}}^{\prime-})-f(E_{\bm{k}}^{\prime+})\right\}}{\sqrt{(\epsilon_a+\epsilon_b)^2+4|\Gamma_{\bm{k}}^{\prime}|^2}}
\label{c25},\\
K^{\prime}=&\dfrac{1}{2N}\sum_{\bm{k}}\dfrac{4s_{\bm{k}}^2tI_0\left\{f(E_{\bm{k}}^{\prime-})-f(E_{\bm{k}}^{\prime+})\right\}}{\sqrt{(\epsilon_a+\epsilon_b)^2+4|\Gamma_{\bm{k}}^{\prime}|^2}} 
\label{c26},
\end{align}
which are real. 
Equations.\;(\ref{c14}) and (\ref{c22}) are then rewritten as:
\begin{align}
\Delta(1-R)=&(\Delta^{\prime}-\Delta^{\prime\dagger})K
\label{c27}, \\
\Delta^{\prime}(1-R^{\prime})=&(\Delta-\Delta^{\dagger})K^{\prime}
\label{c28}.
\end{align}
Taking the Hermite conjugate for Eqs.\;(\ref{c27}) and (\ref{c28}), we obtain: 
\begin{align}
(1-R)(\Delta+\Delta^{\dagger})=&0
\label{c29},\\
(1-R^{\prime})(\Delta^{\prime}+\Delta^{\prime\dagger})=&0
\label{c30}. 
\end{align}
Then, from Eqs.\;(\ref{c29}) and (\ref{c30}), we obtain: 
\begin{align}
\Delta=-\Delta^{\dagger} = i\Delta_0
\label{c31},\\
\Delta^{\prime}=-\Delta^{\prime\dagger}=i\Delta_0^{\prime}
\label{c32}.
\end{align}
Substituting Eqs.\;(\ref{c31}) and (\ref{c32}) into Eqs.\;(\ref{53}) and (\ref{54}), we obtain:
\begin{align}
|\Gamma_{\bm{k}}|^2-|\Gamma_{\bm{k}}^{\prime}|^2=s_{\bm{k}}^2\left\{(V^{\prime\prime})^2-(8tI_0)^2\right\}(\Delta_0^{2}-\Delta_0^{\prime2})
\label{c33}.
\end{align}
If we assume that ${\Delta_0^{\prime}}^2=\Delta_0^2$, then from Eq.\;(\ref{c33}), we obtain:
\begin{align}
|\Gamma_{\bm{k}}^{\prime}|^2=|\Gamma_{\bm{k}}|^2
\label{c34}.
\end{align}
Then, from Eqs.\;(\ref{62}) and (\ref{63}), we obtain 
\begin{align}
E_{\bm{k}}^{\prime\pm}=&-E_{\bm{k}}^{\pm}
\label{c35},\\
f(E_{\bm{k}}^{\prime-})-f(E_{\bm{k}}^{\prime+})=&f(E_{\bm{k}}^{-})-f(E_{\bm{k}}^{+})
\label{c36},
\end{align}
and substituting Eqs.\;(\ref{c35}) and (\ref{c36}) into Eqs.\;(\ref{c23})--(\ref{c26}), we obtain:
\begin{align}
R=&R^{\prime}
\label{c37},\\
K=&K^{\prime}
\label{c38}.
\end{align}
Finally, we obtain two types of self-consistent solutions, $\Delta_0^{\prime}=\pm\Delta_0$, and obtain:
\begin{align}
\Delta=&\Delta^{\prime}\equiv i\Delta_{\rm{dw}}
\label{c39},\\
\Delta=&-\Delta^{\prime} \equiv i\tilde{\Delta}_{\rm{dw}}
\label{c40}. 
\end{align}

\renewcommand{\theequation}{D.\arabic{equation}}
\setcounter{equation}{0}
\section{Calculation details for $\Delta_{\rm{ds}}=\Delta_{\rm{ds}}^{\prime}$ in the $d_{x^2-y^2}$-wave superconducting phase}

We consider the $\Delta_{\rm{ds}}=\Delta_{\rm{ds}}^{\prime}$ case. 
From Eqs.\;(\ref{97}) and (\ref{98}), we obtain: 
\begin{align}
\Lambda=\Lambda^{\prime\dagger}=(V^{\prime}+4tI_0)\Delta_{\rm{ds}}
\label{d1}.
\end{align}
Substituting $m=0$, Eqs.\;(\ref{101})--(\ref{104}), and (\ref{d1}) into Eq.\;(\ref{100}), the total mean-field DSC Hamiltonian can then be written as:
\begin{align}
\mathcal{H}_{\rm{DSC}}^{\rm{MF}}=&\sum_{\bm{k}}\Bigl\{\mathcal{E}_{\bm{k}}^+\bigl(A_{\bm{k}}^{\dagger}A_{\bm{k}}+C_{\bm{k}}^{\dagger}C_{\bm{k}}\bigr)+\mathcal{E}_{\bm{k}}^-\bigl(B_{\bm{k}}^{\dagger}B_{\bm{k}}+D_{\bm{k}}^{\dagger}D_{\bm{k}}\bigr) \notag \\
&\quad+s_{\bm{k}}\Lambda\bigl(D_{\bm{k}}^{\dagger}C_{\bm{k}}+A_{\bm{k}}B_{\bm{k}}^{\dagger}+C_{\bm{k}}^{\dagger}D_{\bm{k}}+B_{\bm{k}}A_{\bm{k}}^{\dagger}\bigr)\Bigr\}
\label{d2},
\end{align}
where $\mathcal{E}_{\bm{k}}^{\pm}$ is defined by Eq.\;(\ref{109}) and we treat $\Lambda$ as a real number. 
The canonical transformation is written as:
\begin{align}
A_{\bm{k}}=&\cos \phi_{\bm{k}} \alpha_{\bm{k}}-\sin \phi_{\bm{k}}\tilde{\alpha}_{\bm{k}}
\label{d3},\\
B_{\bm{k}}=&\sin \phi_{\bm{k}} \alpha_{\bm{k}}+\cos \phi_{\bm{k}}\tilde{\alpha}_{\bm{k}}
\label{d4}, \\
D_{\bm{k}}=&\cos \phi_{\bm{k}} \tilde{\beta}_{\bm{k}}-\sin\phi_{\bm{k}}\beta_{\bm{k}}
\label{d5}, \\
C_{\bm{k}}=&\sin \phi_{\bm{k}} \tilde{\beta}_{\bm{k}}+\cos\phi_{\bm{k}}\beta_{\bm{k}}
\label{d6},
\end{align}
where
\begin{align}
\cos 2\phi_{\bm{k}}=&\dfrac{\epsilon_a}{\sqrt{\epsilon_a^2+s_{\bm{k}}^2\Lambda^2}}
\label{d7},\\
\sin 2 \phi_{\bm{k}}=&\dfrac{-s_{\bm{k}}\Lambda}{\sqrt{\epsilon_a^2+s_{\bm{k}}^2\Lambda^2}}
\label{d8}.
\end{align}
Substituting these transformation into Eq.\;(\ref{d2}), we obtain the diagonal mean-field Hamiltonian:
\begin{align}
\mathcal{H}_{\rm{DSC}}^{\rm{MF}}=&\sum_{\bm{k}}\Bigl\{\xi_{\bm{k}}^+\bigl(\alpha_{\bm{k}}^{\dagger}\alpha_{\bm{k}}+\beta_{\bm{k}}^{\dagger}\beta_{\bm{k}}\bigr)+\xi_{\bm{k}}^-\bigl(\tilde{\alpha}_{\bm{k}}^{\dagger}\tilde{\alpha}_{\bm{k}}+\tilde{\beta}_{\bm{k}}^{\dagger}\tilde{\beta}_{\bm{k}}\bigr)\Bigr\}
\label{d9},
\end{align}
where
\begin{align}
\mathcal{\xi}_{\bm{k}}^{\pm}=&\sqrt{t^2\gamma_{\bm{k}}^2+v_{\rm{dw}}^2\Delta_{\rm{dw}}^2s_{\bm{k}}^2} \pm \sqrt{\epsilon_a^2+s_{\bm{k}}^2\Lambda^2}
\label{d10}.
\end{align}
Note that, from Eqs.\;(\ref{105}) and (\ref{106}), our assumption $\Delta_{\rm{ds}}=\Delta_{\rm{ds}}^{\prime}$ is equivalent to the relation $\langle A_{\bm{k}}^{\dagger}C_{\bm{k}}^{\dagger}\rangle=-\langle D_{\bm{k}}B_{\bm{k}}\rangle$. 
However, from Eqs.\;(\ref{110})--(\ref{113}) and (\ref{d9}), we find that $\langle A_{\bm{k}}^{\dagger}C_{\bm{k}}^{\dagger}\rangle=\langle D_{\bm{k}}B_{\bm{k}}\rangle=0$, which satisfies the aforementioned relation. This result means that $\Delta_{\rm{ds}}=\Delta_{\rm{ds}}^{\prime}$ is self-consistently satisfied. 

Substituting Eq.\;(\ref{d3})--(\ref{d6}) into Eq.\;(\ref{105}), the self-consistent equation for $\Delta_{\rm{ds}}$ is given as: 
\begin{align}
\Delta_{\rm{ds}}=&-\dfrac{\Delta_{\rm{ds}}(V^{\prime}+4tI_0)}{4N}\sum_{\bm{k}}\dfrac{s_{\bm{k}}^2}{
\sqrt{\epsilon_a^2+s_{\bm{k}}^2\Lambda^2}}\bigl\{f(\xi_{\bm{k}}^{-})-f(\xi_{\bm{k}}^{+})\bigr\}
\label{d11}. 
\end{align}

Because $f(\xi_{\bm{k}}^{-})>f(\xi_{\bm{k}}^{+})$ and the integrand of Eq.\;(\ref{d11}) is always positive, $V^{\prime}+4t I_0<0$ must hold to have a nonzero solution of Eq.\;(\ref{d11}). This region is given by: 
\begin{align}
U<\dfrac{1}{2}\bigl(V+\sqrt{V^2+48t^2}\bigr)
\label{d12}.
\end{align}

\renewcommand{\theequation}{E.\arabic{equation}}
\setcounter{equation}{0}
\section{Calculation details for the density of states}

Defining $k_x$ to satisfy $\omega-\epsilon_{\bm{k}}=0$ as $k_x^0$, we obtain: 
\begin{align}
\cos k_x^0(\pm)=&-\frac{1-p^2}{1+p^2}\cos k_y\pm\frac{2p}{1+p^2}\sqrt{\frac{1+p^2}{16p^2}(\omega+\epsilon_a)^2-\cos^2k_y}
\label{e1}.
\end{align}
Because we have
\begin{align}
0\le\omega+\epsilon_a=2\sqrt{(\cos k_x^0+\cos k_y)^2+p^2(\cos k_x^0-\cos k_y)^2}\le4
\label{e2},
\end{align}
the region of $\omega$ is written by $-\epsilon_a\le\omega\le4-\epsilon_a$. 
Therefore, Eq.\;(\ref{130}) is written as: 
\begin{align}
g(\omega)=&\frac{1}{\pi^2}\int_{0}^{\pi}dk_y\sum_{\alpha=\pm}\frac{1}{\bigl|\frac{\partial \epsilon_{\bm{k}}}{\partial k_x}\bigr|_{k_x=k_x^0(\alpha)}}
 \notag \\
		=&\frac{\omega+\epsilon_a}{4\pi^2}\int_{0}^{\pi}dk_y\sum_{\alpha=\pm}\frac{1}{\sin k_x^0(\alpha)\bigl|(1+p^2)\cos k_x^0(\alpha)+(1-p^2)\cos k_y\bigr|}
\label{e3}. 
\end{align}
Furthermore, substituting Eq.\;(\ref{e1}) into Eq.\;(\ref{e3}), we obtain: 
\begin{align}
g(\omega)=&\frac{\omega+\epsilon_a}{8\pi^2p}\int_{0}^{\pi}dk_y\sum_{\alpha=\pm}\frac{1}{\sin k_x^0(\alpha)\sqrt{\frac{1+p^2}{16p^2}(\omega+\epsilon_a)^2-\cos^2k_y}}
\label{e4}. 
\end{align}

\bibliography{your-bib-file}

\end{document}